\documentclass[a4paper,12pt]{article}
\pdfoutput=1
\usepackage{url}

\RequirePackage{color}
\usepackage{graphicx}

\usepackage{amssymb,amsmath}

\usepackage[colon,authoryear]{natbib}
\usepackage{float}

\usepackage{multirow}
\usepackage{booktabs}

\newlength{\verticalcompensationlength}
\newcounter{verticalcompensationrows}
\newcommand{\verticalcompensation}[1]{%
\setcounter{verticalcompensationrows}{#1}%
\addtocounter{verticalcompensationrows}{-1}%
\vspace*{-\value{verticalcompensationrows}\verticalcompensationlength}%
}
\newcommand{\multirowbt}[3]{%
\multirow{#1}{#2}{\verticalcompensation{#1}#3}%
}

\begin{document}

\title{Resonant periodic orbits in the exoplanetary systems}

\author{K. I. Antoniadou, G. Voyatzis\\
Department of Physics, Aristotle University of Thessaloniki, \\54124, Thessaloniki, Greece 
\\kyant@auth.gr, voyatzis@auth.gr}
\maketitle


The final publication is available at springerlink.com\\
http://link.springer.com/article/10.1007/s10509-013-1679-8

\begin{abstract} 
The planetary dynamics of $4/3$, $3/2$, $5/2$, $3/1$ and $4/1$ mean motion resonances is studied by using the model of the general three body problem in a rotating frame and by determining families of periodic orbits for each resonance. Both planar and spatial cases are examined. In the spatial problem, families of periodic orbits are obtained after analytical continuation of vertical critical orbits. The linear stability of orbits is also examined. Concerning initial conditions nearby stable periodic orbits, we obtain long-term planetary stability, while unstable orbits are associated with chaotic evolution that destabilizes the planetary system. Stable periodic orbits are of particular importance in planetary dynamics, since they can host real planetary systems. We found stable orbits up to $60^\circ$ of mutual planetary inclination, but in most families, the stability does not exceed $20^\circ$-$30^\circ$, depending on the planetary mass ratio. Most of these orbits are very eccentric. Stable inclined circular orbits or orbits of low eccentricity were found in the $4/3$ and $5/2$ resonance, respectively. 
\end{abstract}

{\bf keywords} Extrasolar planets, general three body problem, mean motion resonances, periodic orbits, planetary systems.

\section{Introduction}
Over the last decades, there has been a tremendous increase of extrasolar planetary systems discoveries. Many of such systems consist of more than one planet and the study of planetary orbits concerning their long term stability is very interesting. Also, many planets  seem to be locked in mean motion resonance (MMR), with the majority of which in $2/1$ and by descending order, in $3/2$, $5/2$, $3/1$, $4/1$, $4/3$, $5/1$ and $7/2$. However, the present estimation of their initial conditions may change significantly after obtaining additional observational data in the future.  

Our study refers to the dynamics of a two-planet system locked in a MMR. \citet{mbf06} had studied the dynamics of many MMR in the planar case by extracting an appropriate averaged Hamiltonian and computing the families of its stationary points.  Modeling a two-planet system with the general three body problem (GTBP), we can study the dynamics of the non-averaged system by computing families of periodic orbits in a suitable rotating frame \citep{voyhadj05,hadj06,voyatzis08,vkh09}. These families of periodic orbits should coincide with the families of stationary points, provided that the averaged Hamiltonian is sufficiently correct. Also, it has been shown that families of periodic orbits can constitute paths that can drive the migration process and, consequently, trap the planets in a MMR \citep{leepeal02,mebeaumich03,hadjvoy10}. 

All of the above mentioned studies refer to the planar problem. In this paper, we, also, present new results for the planar case but we, mainly, focus on the dynamics of planetary orbits in space. The spatial GTBP, where planets have a mutual inclination, has been studied only for the $2/1$ resonance in \citet{av12}. We herewith determine families of symmetric periodic orbits in all possible configurations of the above mentioned resonances. We compute the spatial families by analytic continuation of vertical critical orbits (v.c.o.) of the planar general problem; a method introduced by \citet{hen} for the restricted problem and extended to the general one by \citet{ichmich80}. Furthermore, we attempt to connect the linear stability of the periodic orbits with the long-term stability of planetary systems close to them.

The paper is organized as follows: In Sect. 2, we briefly present our model and the fundamental concepts of periodic orbits. In Sect. 3, we present the planar families of periodic orbits for different planetary mass ratios along with their horizontal stability and vertical critical orbits. In Sect. 4, we show the generated families of periodic orbits in space, with emphasis given on stable ones. In Sect. 5, we study the long term dynamical evolution of orbits in the vicinity of periodic ones and conclude in Sect. 6. 

\section{The three dimensional GTBP, periodic orbits and stability}
\subsection{Equations of motion}
\begin{figure}[h]
\begin{center}
\includegraphics[width=5cm]{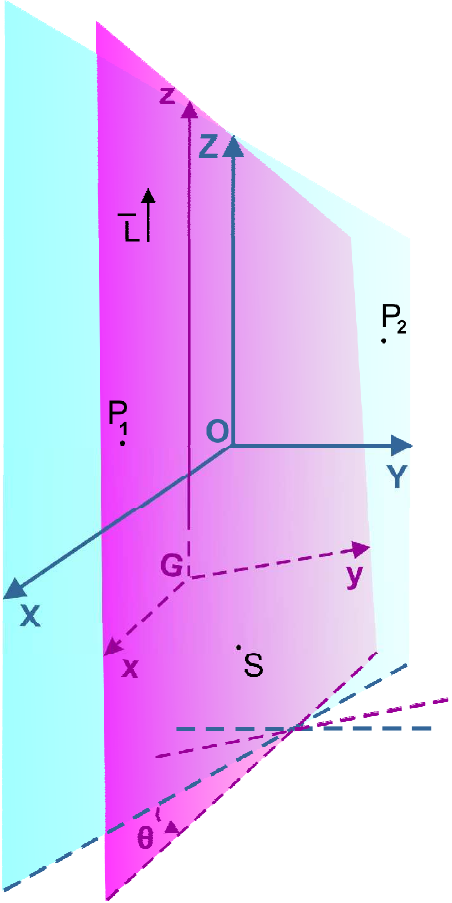}  
\end{center}
\caption{Inertial $OXYZ$ (blue) and rotating $Gxyz$ frame (magenta).}
\label{model}
\end{figure}

We introduce a three dimensional system that consists of a Star, $S$, of mass $m_0$ and two inclined planets, $P_1$ and $P_2$, of masses $m_1$ and $m_2$, respectively, which are considered as point masses. The three bodies move in space $OXYZ$ (inertial frame) under their mutual gravitational attraction, where the origin $O$ is their fixed center of mass and its $Z$-axis is parallel to the constant angular momentum vector, ${\bf L}$, of the system. The system is described by six degrees of freedom, which can be reduced by introducing a suitable rotating frame of reference, $Gxyz$, (Fig. \ref{model}) described in \citet{mich79}, such that:
\begin{enumerate} 
	\item Its origin coincides with the center of mass $G$ of the bodies $S$ and $P_1$.
	\item Its $z$-axis is always parallel to the $Z$-axis.
	\item $S$ and $P_1$ move always on $xz$-plane.
\end{enumerate}
The Lagrangian of the system in the rotating frame of reference is:
\begin{equation}
\begin{array}{l}
\mathfrak{L}=\frac{\displaystyle 1}{\displaystyle 2} \mu[a(\dot x_1^2+\dot z_1^2+x_1^2\dot \theta^2)+\displaystyle b [(\dot x_2^2+\dot y_2^2+\dot z_2^2)\\[0.2cm] 
\quad +\dot\theta^2(x_2^2+y_2^2)+2\dot\theta(x_2\dot y_2-\dot x_2y_2)]]-V,
\label{Lagrangian}
\end{array}
\end{equation}
where
\begin{equation}
\begin{array}{l}
V=-\frac{\displaystyle m_0 m_1}{\displaystyle r_{01}}-\frac{\displaystyle m_0 m_2}{\displaystyle r_{02}}-\frac{\displaystyle m_1 m_2}{\displaystyle r_{12}},\nonumber
\end{array}
\end{equation}
\begin{equation}
\begin{array}{lll}
a=m_1/m_0,&b=m_2/m,&\mu=m_0 + m_1\nonumber
\end{array}
\end{equation}
and
\begin{equation}
\begin{array}{l}
\displaystyle r_{01}^2=(\displaystyle 1+\displaystyle a)^2(\displaystyle x_1^2+z_1^2) , \\[0.3cm] \nonumber
\displaystyle r_{02}^2=(\displaystyle a x_1 +\displaystyle x_2)^2+y_2^2+(\displaystyle a  z_1+z_2)^2, \\[0.3cm] 
\displaystyle r_{12}^2=(\displaystyle x_1 -\displaystyle x_2)^2+y_2^2+ (z_1-z_2)^2.
\label{Lr}
\end{array}
\end{equation}

The Lagrangian of the system does not contain the angle between the axes $OX$ and $Gx$, $\theta$, therefore, the angular momentum $p_\theta=\partial{\mathfrak{L}}/\partial{\dot\theta}$ is constant and given by
\begin{equation}
p_\theta=\mu[ax_1^2\dot \theta+\displaystyle b[\dot \theta (x_2^2+y_2^2)+(x_2\dot y_2-\dot x_2y_2)]]=\textnormal{const.}
\label{Lz}
\end{equation}

Due to the choice of the inertial system, concerning the components of the angular momentum ${\bf L}$, it always holds:
\begin{equation}
\begin{array}{l}
L_{X}=\mu[b(y_2 \dot z_2 -\dot y_2 z_2)-\dot \theta (a x_1 z_1 +b x_2 z_2)]=0,\\[0.3cm] 
L_{Y}=\mu[a(\dot x_{1}z_{1}-x_{1}\dot z_{1})+b(\dot x_{2}z_{2}-x_{2}\dot z_{2})\\[0.3cm] 
\quad\;\; -b\dot \theta y_{2} z_{2}]=0,\\[0.3cm]
L_{Z}=p_\theta.
\label{Lxyz}
\end{array}
\end{equation}
The above restrictions yield
 \begin{equation}
\begin{array}{l}
 z_1=\frac{b}{a} \left ( \frac{y_2\dot z_2-\dot y_2 z_2}{x_1 \dot \theta}-\frac{x_2z_2}{x_1} \right ), \\[10pt] 
\dot z_1=\frac{b}{a x_1}(\dot x_2 z_2-x_2\dot z_2 -\dot{\theta}y_2z_2)+\frac{\dot x_1}{x_1} z_1 
\label{Lxy}
\end{array}
\end{equation}
and
\begin{equation}
\dot \theta=\frac{\frac{p_\theta}{\mu}-\displaystyle b (x_2\dot y_2-\dot x_2 y_2)}{\displaystyle a x_1^2+\displaystyle b (x_2^2+y_2^2)}.
\label{dth}
\end{equation}
Thus, the independent variables in the rotating frame are only $x_1$ for the planet $P_1$ and $x_2$, $y_2$ and $z_2$ for the planet $P_2$ and as a result, the system has been reduced to four degrees of freedom. The equations of motion are:
\begin{equation}
\begin{array}{l}
\ddot x_1=-\frac{m_0 m_2 (x_1-x_2)}{\mu r_{12}^3}-\frac{m_0 m_2 ( a x_1+x_2)}{\mu r_{02}^3}-\frac{\mu x_1}{r_{01}^3}+x_1\dot{\theta}^2\\[0.3cm] 
\ddot x_2=\frac{m m_1 (x_1-x_2)}{\mu r_{12}^3}-\frac{m m_0 (a x_1+x_2)}{\mu r_{02}^3}+x_2\dot{\theta}^2+2 \dot y_2 \dot{\theta}+y_2\ddot{\theta}\\[0.3cm]
\ddot y_2=-\frac{m m_1 y_2}{\mu r_{12}^3}-\frac{m m_0 y_2}{\mu r_{02}^3}+y_2\dot{\theta}^2-2 \dot x_2 \dot{\theta}-x_2\ddot{\theta}\\[0.3cm]
\ddot z_2=\frac{m m_1 (z_1-z_2)}{\mu r_{12}^3}-\frac{m m_0 (a z_1+z_2)}{\mu r_{02}^3}
\label{eq}
\end{array}
\end{equation}

The quantity $\ddot\theta$ is found by differentiating Eq. \eqref{dth} with respect to time. We obtain
\begin{equation}
\ddot \theta=\frac{m_0 m_2}{\mu}\frac{y_2}{x_1}(r^{-3}_{12}-r^{-3}_{02})-2\dot\theta\frac{\dot x_1}{x_1}.
\label{ddth}
\end{equation}

\subsection{Periodic orbits and analytic continuation in space}
We consider the rotating frame $Gxyz$ and define the Poincar\'e section plane $\Pi=\{y_2=0,\dot y_2>0\}$. Then, the periodic orbits are defined as fixed or periodic points of this plane and provided that $y_2(0)=y_2(T)$, where $T$ is the period, they satisfy the conditions:
\begin{equation} \label{apocon}
\begin{array}{l}
\textbf{q}(T)=\textbf{q}_0
\end{array}
\end{equation}
where $\textbf{q}=\left\{x_1,x_2,z_2,\dot x_1,\dot x_2,\dot y_2, \dot z_2\right\}$ and $\textbf{q}_0=\textbf{q}(0)$. We note that for the planar problem is $z_2=\dot{z}_2=0$, which implies that $z_1=\dot{z}_1=0$, too. Due to the existence of the Jacobi integral, one of the above conditions is always fulfilled, when the rest ones are fulfilled. 

If a periodic orbit has two perpendicular crossings with $\Pi$, it is \textit{symmetric}.
Next, we consider the symmetries with respect to the $xz$-plane and the $x$-axis \citep{mich79}.
For a $xz$-symmetric periodic orbit the initial conditions are 
\begin{equation}
\begin{array}{llll}
\{x_{10},x_{20},z_{20},\dot{y}_{20}\} \:\:\text{and}\:\: \dot x_{10}=\dot x_{20}=\dot z_{20}=0.
\label{xzsym}
\end{array}
\end{equation}
For a $x$-symmetric periodic orbit the initial conditions are 
\begin{equation}
\begin{array}{llll}
\{x_{10},x_{20},\dot{y}_{20},\dot{z}_{20}\} \:\:\text{and}\:\: \dot x_{10}=\dot x_{20}=z_{20}=0.
\label{xsym}
\end{array}
\end{equation}

Assuming a planar periodic orbit, we linearize the last of the Eq. \eqref{eq} considering $z_{20}=\zeta_{10}$, $\dot z_{20}=\zeta_{20}$, with $\left|\zeta_{i0}\ll1\right|$ and we obtain the linear system \citep{ichkm78} 
\begin{equation}
\dot \zeta_1=\zeta_2,\quad \dot \zeta_2=A \zeta_1 +B \zeta_2
\label{z22}
\end{equation}
where
\begin{equation}
\begin{array}{l}
A=-\frac{m m_0}{\mu}[(1-\gamma) d^{-3}_{02}+(a+\gamma)d_{12}^{-3}]\\[0.3cm]
B=-\frac{m_0 m_2 y_2}{\mu x_1 \dot\theta}(d^{-3}_{02}-d_{12}^{-3})
\label{AB}
\end{array}
\end{equation}
with 
$$
d^2_{12}=(x_1-x_2)^2+y_2^2,\quad d^2_{02}=(a x_1+x_2)^2+y_2^2
$$ 
and $\gamma=b\frac{x_2 \dot\theta + \dot y_2}{x_1 \dot\theta}$.

If $\Delta(T)=\{\xi_{ij}\}$, $i,j=1,2$, is the monodromy matrix of Eq. \eqref{z22} for a planar periodic orbit of period $T$, we can define the {\em vertical stability index} \citep{hen}
\begin{equation}
a_v=\frac{1}{2}(\xi_{11}+\xi_{22})
\label{av2}
\end{equation}
If $|a_v|<1$ or $|a_v|>1$ the orbit is vertical stable or unstable, respectively. 
Any periodic orbit of the planar problem with $|a_v|=1$ is {\em vertical critical}. A \textit{vertical critical orbit} (v.c.o.) can be analytically continued in space. It is continued by varying the value of $z_2$ for the $xz$-symmetry or $\dot z_2$ for the $x$-symmetry \citep{ichkm78}.

\subsection{Stability of $3D$ periodic orbits} \label{SecStab3D}
\begin{figure*}[!ht]
\begin{center}
\includegraphics[width=12cm]{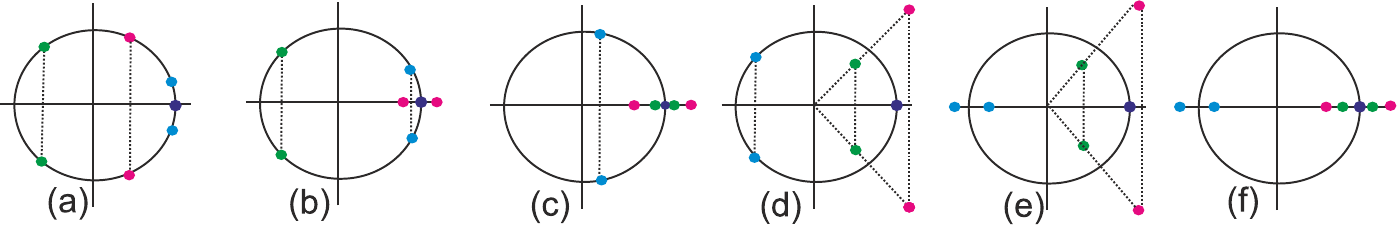}  
\end{center}
\caption{Stability types of $3D$ periodic orbits a) stability b) simple instability c) double instability d) complex instability e) u-complex instability f) triple instability.}
\label{stab}
\end{figure*}

The linear stability of periodic orbits is found by computing the eigenvalues of the $8\times8$ monodromy matrix of the variational equations of the system \eqref{eq}. Since the monodromy matrix is symplectic, we have 4 pairs of conjugate eigenvalues. We compute them by using the LAPACK routines of $Mathematica$. Iff all the eigenvalues lie on the unit circle, then the periodic orbit is \textit{linearly stable}. Due to the existence of the energy integral, one pair of eigenvalues is equal to unity. The location of the other three pairs (see Fig. \ref{stab}), may lead to simple, double, triple, complex or (as we call it here) $u$-complex instability (\citealt{marchal90,sk01}).  

In some cases, due to the limited accuracy, we cannot be sure whether an eigenvalue lies on the unit circle, or not. Therefore, apart from the computation of the eigenvalues we can estimate the evolution stability by using as index the Fast Lyapunov Indicator computed along the periodic orbit (\citealt{froe97,voyatzis08}). 

\subsection{Planetary resonant configurations and presentation of families of periodic orbits} \label{SecPres}
Having assumed that $m_0\gg m_{1,2}$ the periodic orbits in the rotating frame correspond to almost Keplerian ellipses described by the osculating orbital elements $a_i$ (semimajor axis), $e_i$ (eccentricity), $i_i$ (inclination), $\omega_i$ (argument of pericenter), $\Omega_i$ (longitude of ascending node) and $\lambda_i$ (mean longitude). Throughout the present study we take $a_1<a_2$ (thus, subscript 1 corresponds to the inner planet and 2 to the outer). The total mass of the system, $m$, is normalized to unity, i.e. $m=m_0+m_1+m_2=1$. The inner planet\rq{}s mass is fixed to $m_J=10^{-3}$ (Jupiter\rq{}s mass), unless otherwise stated. 


Mean motion resonances are associated with the non-circular families of periodic orbits for which $\frac{n_1}{n_2}\approx\frac{p+q}{p}$, where $q,\,p\neq0$ are integers and $n_i$ denotes the mean motion of the planet $P_i$. In this case, we can introduce the resonant angles  
\begin{equation}
\begin{array}{l}
\theta _{1}=(p+q)\lambda _{1}-p\lambda _{2}-q\varpi _{1}\\ 
\theta _{2}=(p+q)\lambda _{1}-p\lambda _{2}-q\varpi _{2}\\
\end{array}
\end{equation}
Also, we may refer to the apsidal difference $\Delta\varpi=\varpi_1-\varpi_2$. For symmetric periodic orbits, we have $\Delta\varpi=0$ (aligned planets) or $\Delta\varpi=\pi$ (antialigned planets). Also, we assume that at $t=0$ the planets are placed at periastron, $M_i=0$ or apoastron, $M_i=\pi$, where $M_i$ indicates the mean anomaly of planet $P_i$. Thus, we get four resonant configurations and we can use the resonant angles $(\theta_1,\theta_2)$ to distinct them. Therefore, the four configurations arising are $(0,0)$, $(\pi,0)$, $(\pi,\pi)$ and $(0,\pi)$.
However, for resonances with $q=$even, we always have, at $t=0$, $\theta_1-\theta_2=q\Delta\varpi=0$ and, consequently, the pair of angles $(\theta_1,\theta_2)$ would not allow us to distinguish the different configurations. In such cases, e.g. the $3/1$ resonance, where $q=2$, we shall use the pair $(\theta_3,\theta_1)$, where the resonant angle $\theta_3$ is defined as
\begin{equation}
\theta _{3}=(p+q)\lambda _{1}-p\lambda _{2}-\tfrac{q}{2}(\varpi _{1}+\varpi _{2}).
\end{equation}

Following the literature \citep{mbf06}, we project the families of planar periodic orbits in the eccentricity plane $(e_1,e_2)$. In order to distinguish the families of different configurations in the projection plane we use both positive and negative values for the eccentricities. The positive values of $e_i$ correspond to $\theta_i=0$ and the negative ones to $\theta_i=\pi$. 
 
Evidently, the periodic solutions generally depend on planetary masses. It has been shown by \citep{beau03,fer06} that the families of periodic orbits on the projection plane $(e_1,e_2)$ depend on the mass ratio of the planets, $\rho=\frac{m_2}{m_1}$ and not on planetary masses individually, provided that $m_i\ll m_0$. The various families belonging to the same configuration differ from one another in $\rho$, which usually extends in our computations from $0.01$ to $20$. The families of the same resonance and same configuration (but different $\rho$) form a {\em group} of families. 

The v.c.o. of the families are classified according to the symmetry of the periodic orbits they generate: $xz$-symmetric and $x$-symmetric orbits are denoted with symbols $\widehat F$ and $\widehat G$, respectively. They can be, also, classified according to the resonance, $\frac{p+q}{p}$ and the configuration $(\theta_1,\theta_2)$ they belong to\footnote{We remind that for even values of $q$, we consider the pair $(\theta_3,\theta_1)$.}. 

Concerning the spatial problem, we, also, have $\Delta\varpi=0$ or $\pi$ and $\theta_i=0$ or $\pi$. Thus, the spatial families are denoted with symbols $F$ and $G$, similarly to the v.c.o. where they start from. Also, the resonance and the configuration is indicated with a superscript and a subscript, respectively, i.e. we name the families as $F^{p+q/p}_{(\theta_{1},\theta_{2})}$ or $G^{p+q/p}_{(\theta_{1},\theta_{2})}$. In case there exist more than one groups of families or v.c.o. in the same configuration and resonance, symbols $F$ and $G$ are primed. We present the spatial families in the $3D$ projection space ($e_1,e_2,\Delta i)$, where $\Delta i$ is the mutual inclination of the planets given by the cosine rule
\begin{equation}\cos\Delta i=\cos i_1 \cos i_2+\sin i_1 \sin i_2\cos(\Omega_1-\Omega_2).\nonumber\end{equation} 

We have computed the periodic orbits by solving numerically the equations (\ref{eq}) of the rotating frame using the Bulirsch-Stoer integrator and setting the minimum accuracy to $10^{-14}$. The periodicity conditions (\ref{xzsym}) or (\ref{xsym}) are satisfied at least up to 12 decimal digits after successive differential approximations. Since we aim to demonstrate a global view of stability and instability regions in phase space, we do not provide the accurate data of our computations. Instead, we present graphs of eccentricities and mutual inclination to which the estimated observational data can be assigned. The ratio of planetary semimajor axes can be approximated by the particular resonance, i.e. $\frac{a_1}{a_2}\approx \left (\frac{p+q}{p} \right )^{-2/3}$. We should remark that in extrasolar systems with very massive planets (e.g. $m_i\gtrsim 10~m_J$), the mean motion ratio may differ significantly from the rational value $(p+q)/p$.

\section{Planar families of symmetric periodic orbits}
We herewith  present families of symmetric planar periodic orbits in the MMRs 4/3, 3/2, 5/2, 3/1 and 4/1. We have computed the vertical stability of periodic orbits in each family and present the v.c.o. found. Stable v.c.o. are of particular importance, since they can yield, after continuation process, families of spatial stable orbits, where extrasolar planets could be trapped to.
 
We depict the characteristic curves of the families on the projection plane $(e_1,e_2)$. The segments of the families which consist of stable orbits are presented with bold blue lines, while the segments of unstable ones are depicted with red lines. The v.c.o. which generate $xz$-symmetric periodic orbits in space are presented by magenta-coloured dots, whereas the $x$-symmetric with green dots. Bold gray lines represent close encounters between the planets.

\subsection{$4/3$ resonance}\label{P43}
\begin{figure*}[!htp]
\begin{center}
\includegraphics[width=10cm,height=10cm]{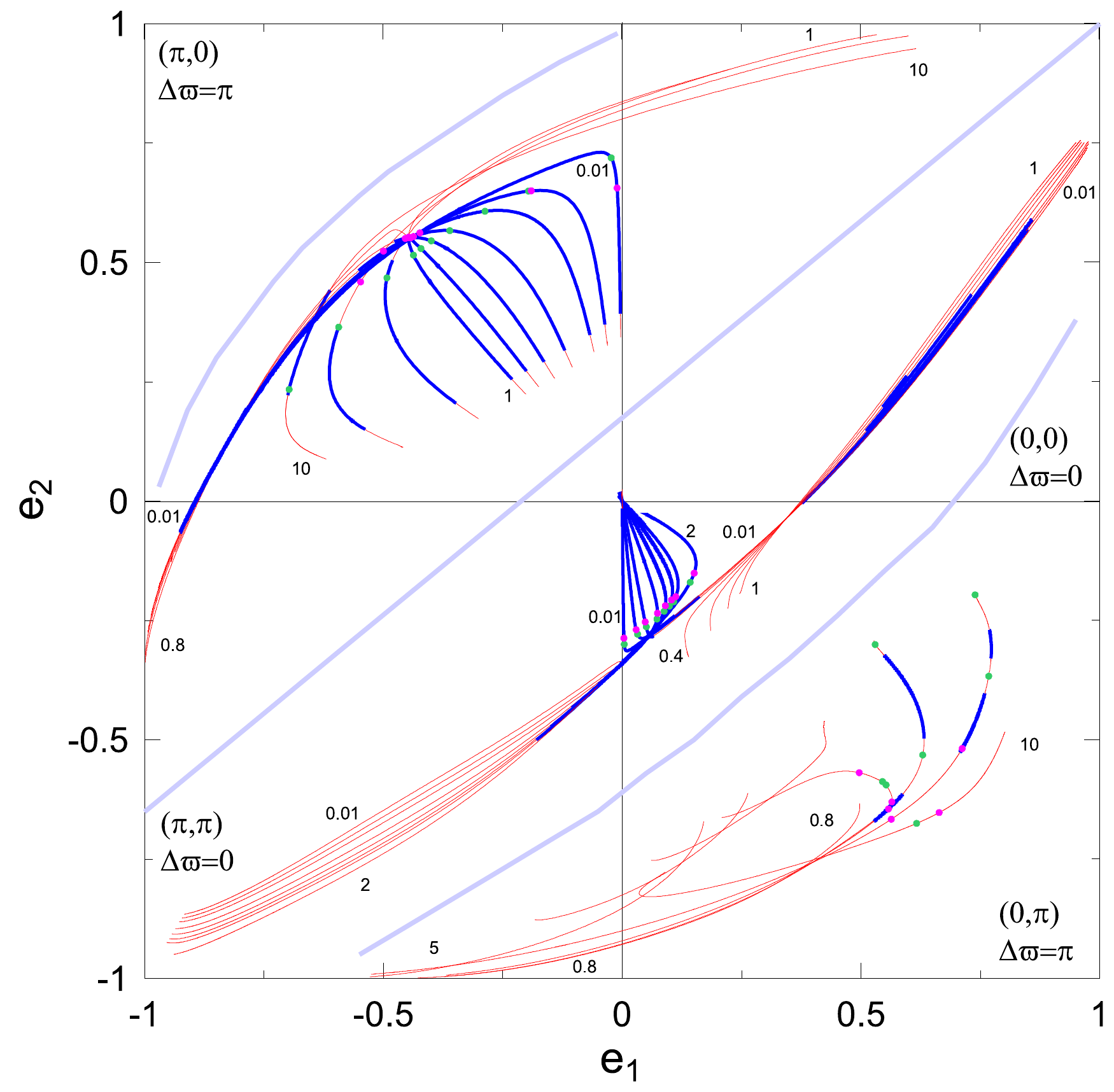}  
\end{center}
\caption{Planar families of symmetric periodic orbits in $4/3$ resonance for various (indicated) mass ratios. The particular configuration $(\theta_1,\theta_2)$ for each quarter is indicated. Bold blue lines represent stable families, while the red ones the unstable families.  Magenta or green coloured dots correspond to $\widehat F$ and $\widehat G$ v.c.o., respectively. The mass ratio $\rho$ for some indicative families is given. Bold gray lines stand for close encounters between the planets.}
\label{43}
\end{figure*}

Extrasolar planetary systems close to $4/3$ mean motion resonance are e.g. HD $200964$ \citep{wit12} and Kepler $29$. In Fig. \ref{43}, we present the planar families of periodic orbits along with their stability and the v.c.o. they possess. There exists one group of families of periodic orbits that bifurcates from $(0,0)$ (circular family) and starts as stable; it belongs to the configuration $(0,\pi)$. In the configuration $(\pi,0)$, we obtain stable orbits above the collision line. This means that the planetary orbits intersect each other. However, due to the phase protection mechanism offered by the resonance, the planets avoid collisions and their orbits are stable. The same holds for the same configuration of all studied resonances. 

In Table \ref{43t}, we have classified the v.c.o. in accordance with the configuration to which they belong and the symmetry of spatial periodic orbits they generate. The configurations $(0,0)$ and $(\pi,\pi)$ do not have any v.c.o.. There are, also, circular v.c.o., at $(e_1,e_2)\approx(0,0)$, which are not included in the Table. They exist for mass ratio values approximately in the interval $0.35<\rho<0.85$, and as we will see in Sect. \ref{S43}, circular families of inclined orbits bifurcate from them. 

\begin{table*}[!htp]
\caption{Eccentricity values of v.c.o. in $4/3$ resonance.}
\begin{tabular}[b]{lllllllll}
\toprule
 & $\widehat{F}^{4/3}_{(0,\pi)}$ & &$\widehat{G}^{4/3}_{(0,\pi)}$ & & $\widehat{F}^{4/3}_{(\pi,0)}$ & & $\widehat{G}^{4/3}_{(\pi,0)}$& \\
 \cmidrule{2-9}
$\rho$ & $e_1$ & $e_2$ & $e_1$ & $e_2$& $e_1$ & $e_2$ & $e_1$ & $e_2$\\
\midrule
\multirowbt{1}{*}{0.01}&0.003&0.287& 0.004&0.299&0.010&0.655& 0.022&0.719  \\
\midrule
\multirowbt{1}{*}{0.1}&0.029&0.269& 0.032&0.278&0.190&0.649 & 0.196&0.649  \\
\midrule
\multirowbt{1}{*}{0.2}&0.048&0.253&0.050& 0.263&0.423&0.561&0.287& 0.607 \\
\midrule
\multirowbt{1}{*}{0.4}&0.073&0.234&0.072 &0.247&0.437&0.554&0.360 &0.566 \\
\midrule
\multirowbt{1}{*}{0.6}&0.090&0.219&0.087&0.230&0.442&0.551 &0.399&0.545\\
\midrule
\multirowbt{1}{*}{0.8}&0.103&0.207&0.098 &0.219 &0.446&0.551&0.421 &0.529 \\
\midrule
\multirowbt{3}{*}{1}&0.112&0.200 & 0.107& 0.210 &0.447&0.553 & 0.437&0.515\\
\cmidrule{2-9}
 & 0.496 & 0.568 & 0.545& 0.587 \\
\cmidrule{2-5}
 & 0.564 & 0.629 &0.552 & 0.594\\
\midrule
\multirowbt{3}{*}{2}&0.150&0.150&0.142 &0.169&0.454&0.550&0.492 &0.467 \\
\cmidrule{2-9}
 & 0.557 & 0.644 &0.629 & 0.531 \\
 \cmidrule{2-5}
 &  &  &0.529 & 0.300 \\
\midrule
\multirowbt{2}{*}{5}&0.563&0.666& 0.767& 0.366&0.547&0.459& 0.593& 0.364 \\
\cmidrule{2-9}
 & 0.711& 0.518 & 0.739 & 0.196 & 0.499& 0.523 &  &  \\
\midrule
\multirowbt{1}{*}{10}&0.663&0.652&0.616 &  0.675&&&0.697 &0.233 \\
\bottomrule\end{tabular}
\label{43t}
\end{table*}

\subsection{$3/2$ resonance}
\begin{figure*}[!htp]
\begin{center}
\includegraphics[width=10cm,height=10cm]{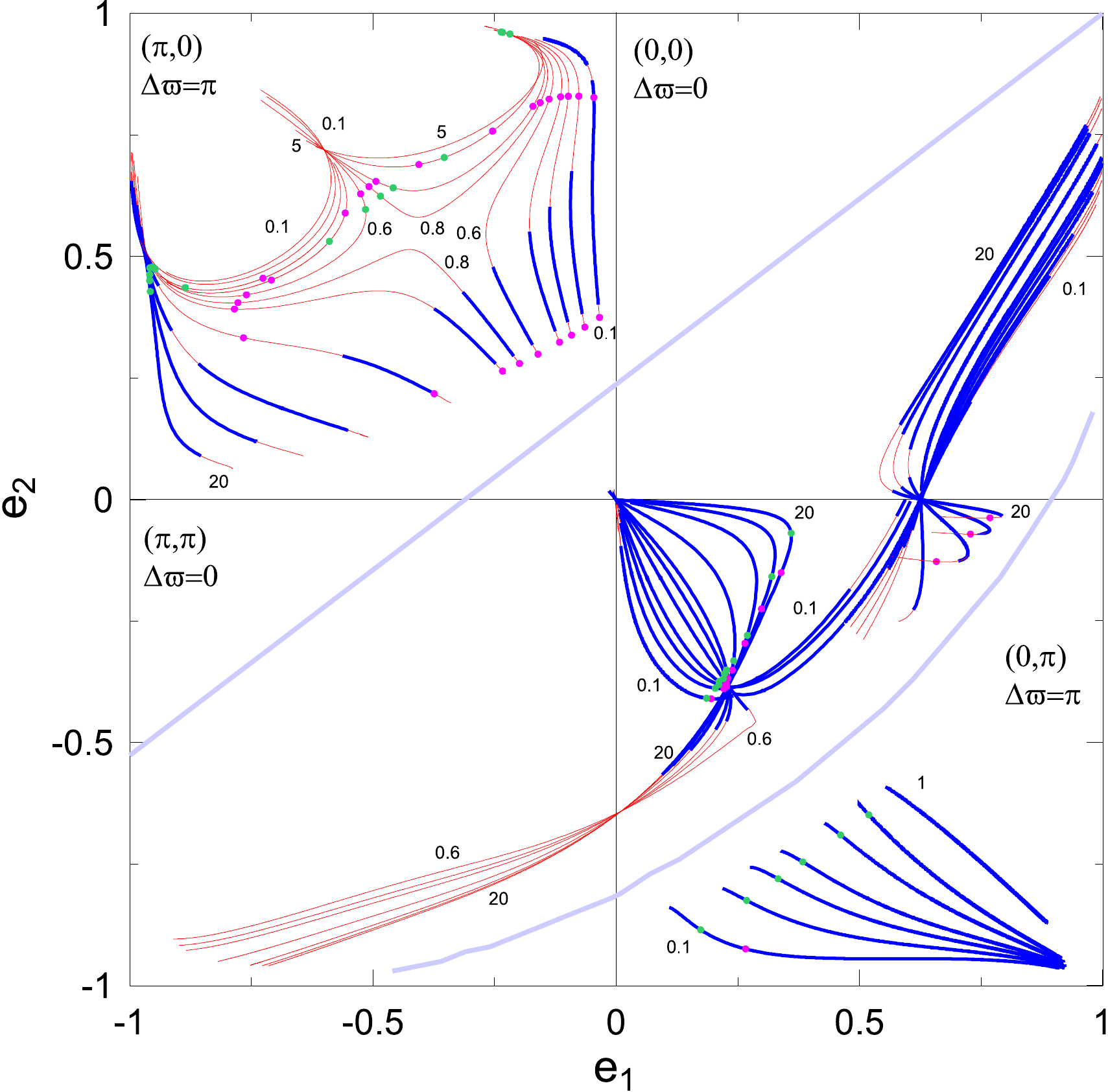} 
\end{center}
\caption{Planar families of symmetric periodic orbits in $3/2$ resonance. They are presented as in Fig. \ref{43}.}
\label{32}
\end{figure*}

The $3/2$ mean motion resonance is apparent in many exosolar planetary systems, e.g. HD $45364$ \citep{rein10}, KOI $55$ \citep{cal12}, and a lot of Kepler systems (see \cite{stef13,stef12}). In Fig. \ref{32}, we present 3/2 resonant families of periodic orbits and the v.c.o. they possess.

In configuration $(0,\pi)$, similarly to the $4/3$ resonance, there is a group of families of stable periodic orbits that bifurcates from the circular family of periodic orbits. The configuration $(\pi,0)$ has, also, families of periodic orbits with stable regions. The v.c.o. of these configurations are shown in Table \ref{32t}. The configuration $(\pi,\pi)$ does not have any v.c.o. and its families are totally unstable. The families in the configuration $(0,0)$ have no v.c.o., too.

\begin{table*}[!htp]
\caption{Eccentricity values of v.c.o. in $3/2$ resonance.}
\scalebox{0.8}{
\begin{tabular}[b]{lllllllll}
\toprule
 & $\widehat{F}^{3/2}_{(0,\pi)}$ & &$\widehat{G}^{3/2}_{(0,\pi)}$ & & $\widehat{F}^{3/2}_{(\pi,0)}$ & & $\widehat{G}^{3/2}_{(\pi,0)}$& \\
\cmidrule{2-9}
$\rho$ & $e_1$ & $e_2$ & $e_1$ & $e_2$& $e_1$ & $e_2$ & $e_1$ & $e_2$\\\midrule
\multirowbt{2}{*}{0.1} & 0.195 & 0.409 & 0.185 & 0.408 & 0.034 & 0.374 &  & \\
\cmidrule{2-7}
 & 0.266 & 0.924 & 0.173 & 0.885 & 0.045 & 0.826 & &\\
\midrule
\multirowbt{2}{*}{0.2} & 0.221 & 0.389 & 0.203 & 0.388 & 0.064 & 0.354 &  &\\
\cmidrule{2-7}
 &  &  & 0.268 & 0.824 & 0.076 & 0.829 & & \\
\midrule
\multirowbt{3}{*}{0.3} & 0.227 & 0.384 & 0.211 & 0.380 & 0.091 & 0.337 & 0.885 & 0.435\\
\cmidrule{2-9}
 & &  & 0.332 & 0.780& 0.726 & 0.455 & & \\
\cmidrule{4-7}
 &&&&& 0.098 & 0.829 & & \\
\midrule
\multirowbt{4}{*}{0.4} & 0.225 & 0.381 & 0.211 & 0.374& 0.116 & 0.323 & 0.589 & 0.530  \\
\cmidrule{2-9}
 &  &  & 0.383 & 0.745& 0.114 & 0.828 & &\\
 \cmidrule{4-7}
 &&&&& 0.708 & 0.451 & & \\
\cmidrule{6-7}
&&&&&0.557 & 0.589 & & \\
\midrule
\multirowbt{4}{*}{0.6} & 0.226 & 0.376 & 0.220 & 0.368 & 0.160 & 0.298 & 0.515 & 0.596\\
\cmidrule{2-9}
 &  &  & 0.461 & 0.690& 0.760 & 0.421 & 0.947 & 0.474\\
 \cmidrule{4-9}
  &&&&& 0.525 & 0.628 & & \\
\cmidrule{6-7}
&&&& & 0.137 & 0.823 & &  \\
\midrule
\multirowbt{4}{*}{0.8} & 0.226 & 0.368 & 0.222 & 0.358 & 0.198 & 0.279 & 0.952 & 0.477\\
\cmidrule{2-9}
 &  &  & 0.519 & 0.648& 0.777 & 0.404 & 0.484 & 0.624 \\
 \cmidrule{4-9}
&&&& & 0.508 & 0.643 & 0.236 & 0.961 \\
\cmidrule{6-9}
&&&& & 0.156 & 0.816 & & \\
\midrule
\multirowbt{4}{*}{1} & 0.229 & 0.367 & 0.226 & 0.350 & 0.233 & 0.264 & 0.954 & 0.477 \\
\cmidrule{2-9}
&&&& & 0.784 & 0.391 & 0.458 & 0.641 \\
\cmidrule{6-9}
&&&& & 0.493 & 0.654 & 0.233 & 0.960 \\
\cmidrule{6-9}
&&&& & 0.171 & 0.808 & & \\
\midrule
\multirowbt{4}{*}{2} & 0.238 & 0.350 & 0.241 & 0.331& 0.374 & 0.217 & 0.958 & 0.476 \\
\cmidrule{2-9}
&&&&  & 0.766 & 0.332 & 0.353 & 0.703 \\
\cmidrule{6-9}
&&&&  & 0.405 & 0.688 & 0.218 & 0.957  \\
\cmidrule{6-9}
&&&& & 0.254 & 0.757 & & \\
\midrule
\multirowbt{2}{*}{5} & 0.264 & 0.295 & 0.269 & 0.279 &  &  & 0.959 & 0.462\\
\cmidrule{2-9}
 & 0.658 & 0.127 &  & &&&&\\
\midrule
\multirowbt{2}{*}{10} & 0.298 & 0.224 & 0.320 & 0.158&  &  & 0.959 & 0.450 \\
\cmidrule{2-9}
 & 0.728 & 0.071 &  & &&&& \\
\midrule
\multirowbt{2}{*}{20} & 0.339 & 0.150 & 0.359 & 0.069&  &  & 0.957 & 0.428 \\
\cmidrule{2-9}
 & 0.768 & 0.037 &  & &&&& \\
\bottomrule\end{tabular}}
\label{32t}
\end{table*}

\subsection{$5/2$ resonance}\label{P52}
\begin{figure*}[!htp]
\begin{center}
\includegraphics[width=10cm,height=10cm]{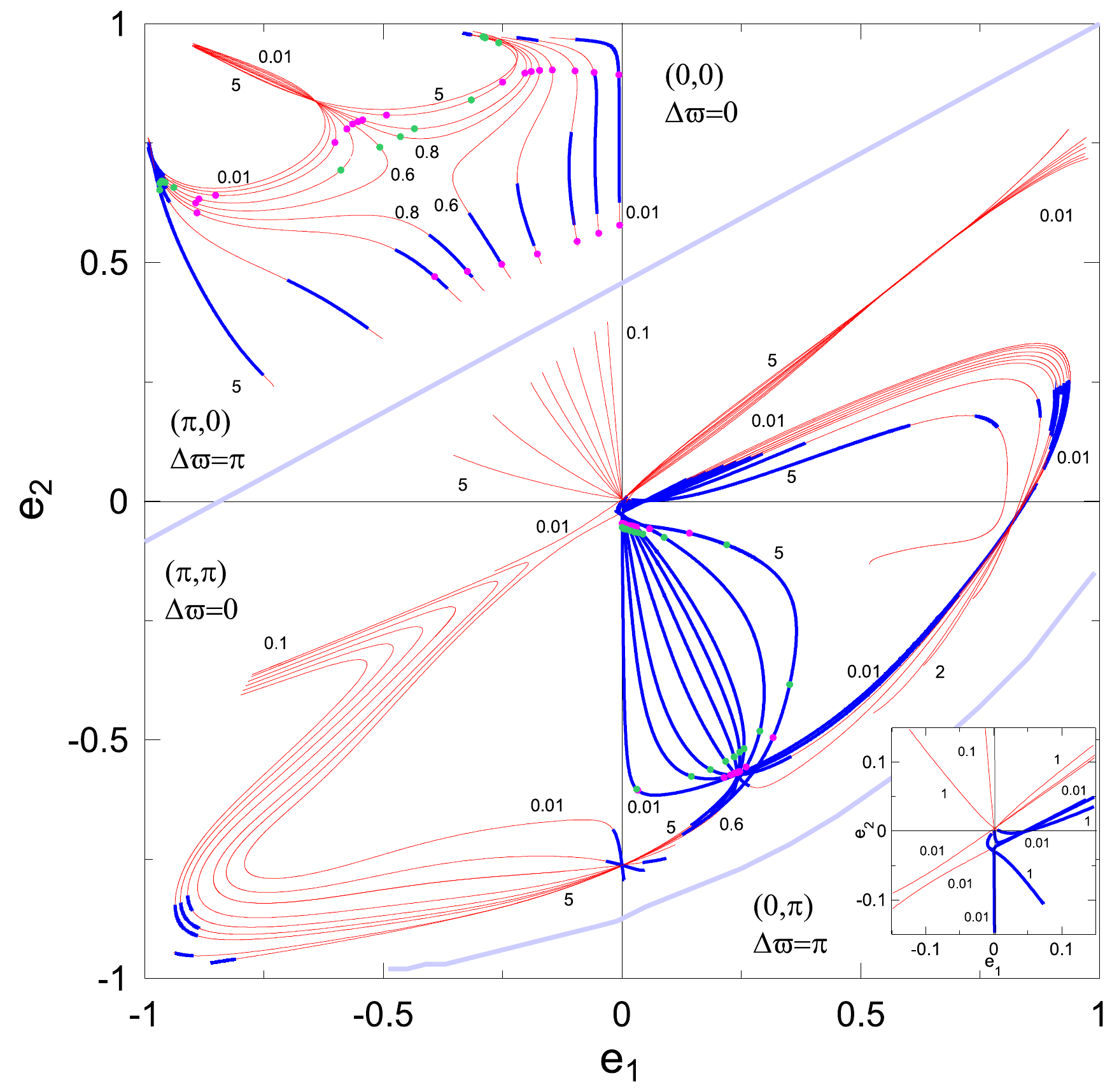}  
\end{center}
\caption{Presentation of planar families of symmetric periodic orbits in $5/2$ resonance, as in Fig. \ref{43}.}
\label{52}
\end{figure*}

The extrasolar planetary systems  HD $1461$ and HD $21693$ are locked near the $5/2$ resonance, while planets of the multiplanetary systems HD $10180$ (f,e) \citep{lovis11} and HD $181433$(d,c) \citep{camp11,camp13} seem to evolve in it. Particular families of periodic orbits of this resonance were given by \cite{psy52}. In Fig. \ref{52}, we present the structure of families of periodic orbits belonging to each configuration.

For each configuration we obtain families that bifurcate from circular orbits. At $(0,0)$ the picture is quite complicated and we added a magnification plot with few representative family curves. Segments of stability along the families are evident in every configuration. In Table \ref{52t}, we show the v.c.o. of the configurations $(0,\pi)$ and $(\pi,0)$. The rest ones do not have any v.c.o.. 

\begin{table*}[!htp]
\caption{Eccentricity values of v.c.o. in $5/2$ resonance.}
\scalebox{0.85}{
\begin{tabular}[b]{lllllllll}
\toprule
 & $\widehat{F}^{5/2}_{(0,\pi)}$ & &$\widehat{G}^{5/2}_{(0,\pi)}$ & & $\widehat{F}^{5/2}_{(\pi,0)}$ & & $\widehat{G}^{5/2}_{(\pi,0)}$& \\
 \cmidrule{2-9}
$\rho$ & $e_1$ & $e_2$ & $e_1$ & $e_2$& $e_1$ & $e_2$ & $e_1$ & $e_2$\\
\midrule
\multirowbt{2}{*}{0.01} & 0.032 & 0.605 & 0.030 & 0.603 & 0.005 & 0.577 &  &\\
\cmidrule{2-7}
 & 0.0003 & 0.046 & 0.0004 & 0.055 & 0.006 & 0.892 & &  \\
\midrule
\multirowbt{2}{*}{0.1} & 0.213 & 0.579 & 0.145 & 0.576& 0.048 & 0.560 &  &   \\
\cmidrule{2-7}
 & 0.003 & 0.047 & 0.004 & 0.058 & 0.058 & 0.897 & & \\
\midrule
\multirowbt{4}{*}{0.2} & 0.227 & 0.573 & 0.184 & 0.562& 0.094 & 0.544 &  &   \\
\cmidrule{2-7}
 & 0.006 & 0.048 & 0.008 & 0.058& 0.851 & 0.640 & & \\
 \cmidrule{2-7}
&&&&  & 0.601 & 0.751 & & \\
\cmidrule{6-7}
&&&& & 0.098 & 0.900 & & \\
\midrule
\multirowbt{4}{*}{0.4} & 0.235 & 0.569 & 0.216 & 0.545& 0.177 & 0.517 & 0.938 & 0.656  \\
\cmidrule{2-9}
& 0.013 & 0.051 & 0.017 & 0.062& 0.886 & 0.632 & 0.588 & 0.693\\
 \cmidrule{2-9}
&&&& & 0.576 & 0.779 & & \\
\cmidrule{6-7}
&&&& & 0.146 & 0.902 & & \\
\midrule
\multirowbt{4}{*}{0.6} & 0.240 & 0.569 & 0.234 & 0.535& 0.252 & 0.495 & 0.956 & 0.667  \\
\cmidrule{2-9}
 & 0.018 & 0.051 & 0.026 & 0.065 & 0.892 & 0.624 & 0.507 & 0.741 \\
\cmidrule{2-9}
&&&& & 0.564 & 0.789 & 0.291 & 0.973 \\
\cmidrule{6-9}
&&&& & 0.172 & 0.902 & & \\
\midrule
\multirowbt{4}{*}{0.8} & 0.243 & 0.567 & 0.246 & 0.525 & 0.892 & 0.614 & 0.962 & 0.669\\
\cmidrule{2-9}
 & 0.024 & 0.052 & 0.034 & 0.066& 0.324 & 0.481 & 0.464 & 0.763 \\
\cmidrule{2-9}
&&&& & 0.552 & 0.794 & 0.289 & 0.971\\
\cmidrule{6-9}
&&&& & 0.190 & 0.899 & & \\
\midrule
\multirowbt{4}{*}{1} & 0.246 & 0.566 & 0.255 & 0.518 & 0.392 & 0.469 & 0.964 & 0.670\\
\cmidrule{2-9}
 & 0.030 & 0.054 & 0.043 & 0.069  & 0.890 & 0.603 & 0.434 & 0.780 \\
\cmidrule{2-9}
&&&& & 0.543 & 0.798 & 0.286 & 0.970 \\
\cmidrule{6-9}
&&&& & 0.203 & 0.896 & & \\
\midrule
\multirowbt{3}{*}{2} & 0.259 & 0.556 & 0.288 & 0.481& 0.493 & 0.808  & 0.967 & 0.665 \\
\cmidrule{2-9}
 & 0.057 & 0.057 & 0.088 & 0.076  & 0.250 & 0.877 & 0.315 & 0.840 \\
\cmidrule{2-9}
&&&& &  &  & 0.258 & 0.960 \\
\midrule
\multirowbt{2}{*}{5} & 0.316 & 0.495 & 0.351 & 0.384 &  &  & 0.968 & 0.651\\
\cmidrule{2-9}
 & 0.141 & 0.067 & 0.219 & 0.091 &&&&\\
\bottomrule\end{tabular}}
\label{52t}
\end{table*}

\subsection{$3/1$ resonance}
\begin{figure*}[!htp]
\begin{center}
\includegraphics[width=10cm,height=10cm]{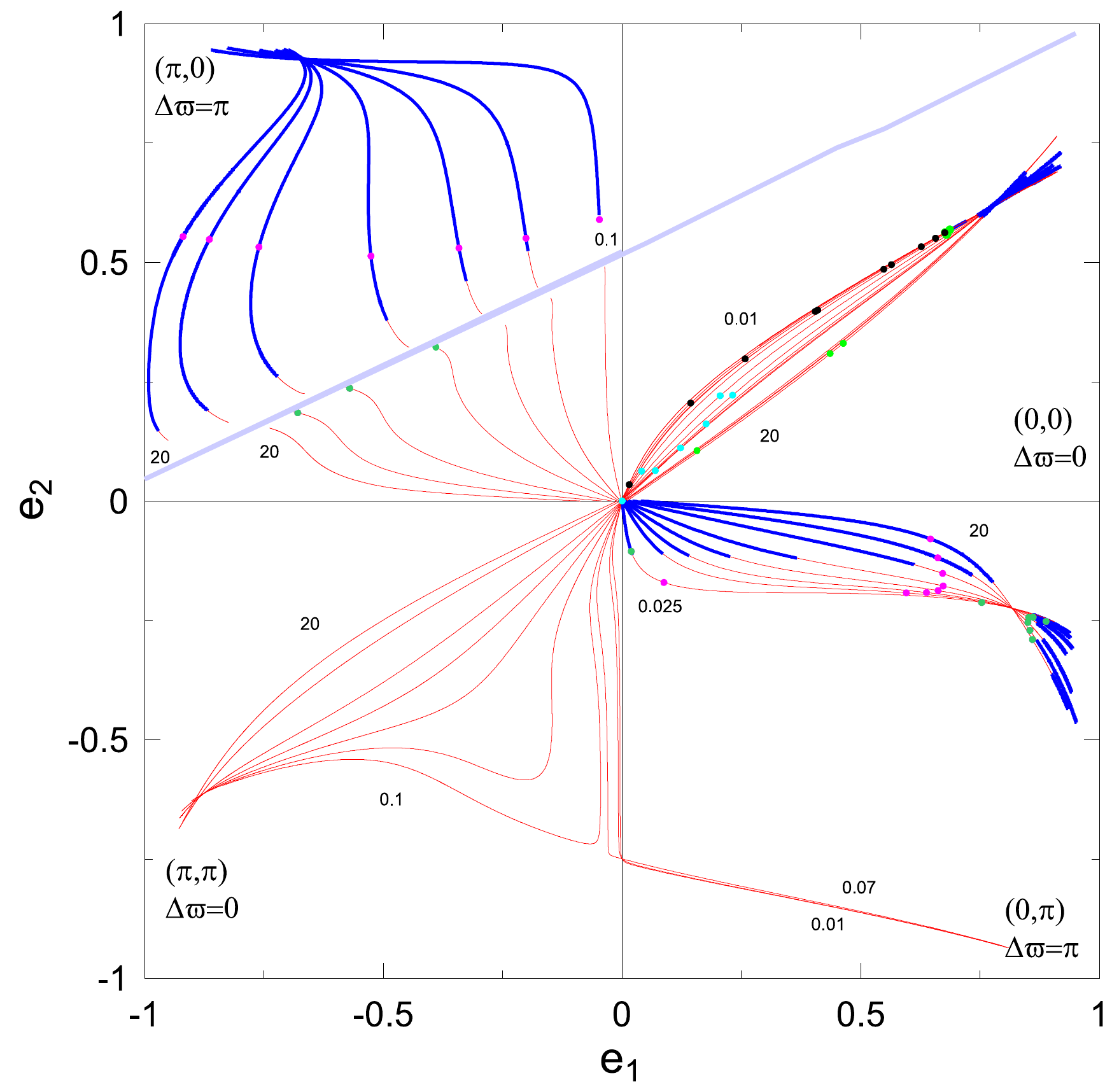} 
\end{center}
\caption{Presentation of planar families of symmetric periodic orbits in $3/1$ resonance, as in Fig. \ref{43}.}
\label{31}
\end{figure*}

The $3/1$ resonance has been observed in the following planetary systems: HD $60532$ \citep{lask09}, HD $10180$(e,d)  \citep{lovis11}, and possibly in HD $20781$ and HD $20003$. The structure and stability of each group of families belonging to different configurations is thoroughly described in \cite{voyatzis08} and following this study, we kept the inner planet\rq{}s mass fixed to $8\,10^{-4}$. Furthermore, in Fig. \ref{31}, we show the v.c.o. of each family. 

In configuration $(0,\pi)$ there are couples of $x$- symmetric v.c.o. from which, as we shall see in Sect. \ref{331}, one family bifurcates forming a bridge from one v.c.o. to the other for mass ratios $\rho<0.158$. V.c.o. of such symmetry appear again for $e_1>0.75$. From the rest v.c.o. $xz$-symmetric periodic orbits bifurcate. We note that this behaviour was, also, reported in the study of $2/1$ resonance (\citet{av12}). Stable v.c.o., also, appear in configuration $(\pi,0)$. All of them are given in Table \ref{31t}.

The configuration $(0,0)$ has many v.c.o. that belong to unstable families of periodic orbits. However, we should state the bifurcations that take place as $\rho$ increases, in order to show the complexity of the dynamics:  i) For $\rho<0.02375$ there exists one $\widehat{G}$ v.c.o. (i.e. of $x$-symmetry) in each family (green dots). ii) For $0.025<\rho<0.105$ there are couples of $\widehat{G}$ v.c.o. (black dots). iii) For $0.11<\rho<0.14375$ there is no v.c.o. iv) For $0.15<\rho<1.875$ there appears a $\widehat{F}$ v.c.o., from which $xz$-symmetric spatial periodic orbits bifurcate (gray dots). v) For $1.9<\rho<13.5$ there are no v.c.o. vi) For $13.5<\rho<20$ there is again one $\widehat{G}$ v.c.o. of consequent $x$-symmetry (green dots).

\begin{table*}[!htp]
\caption{Eccentricity values of v.c.o. in $3/1$ resonance.}
\begin{tabular}[b]{lllllllll}
\toprule
 & $\widehat{F}^{3/1}_{(\pi,0)}$ & &$\widehat{G}^{3/1}_{(\pi,0)}$ & & $\widehat{F}^{3/1}_{(0,\pi)}$ & & $\widehat{G}^{3/1}_{(0,\pi)}$& \\
 \cmidrule{2-9}
$\rho$ & $e_1$ & $e_2$ & $e_1$ & $e_2$& $e_1$ & $e_2$ & $e_1$ & $e_2$\\
\midrule
\multirowbt{2}{*}{0.025}&&&&& 0.087 & 0.170 & 0.019 & 0.105\\
\cmidrule{6-9}
&&&& &  &  & 0.753 & 0.212\\
\midrule
\multirowbt{1}{*}{0.1}&0.047&0.589& &&0.494 & 0.191 & 0.158 & 0.171\\
\cmidrule{6-9}
&&&& &  &  & 0.693 & 0.203\\
\midrule
\multirowbt{1}{*}{0.25}&&&&& 0.595 & 0.192 & & \\
\midrule
\multirowbt{1}{*}{0.5}&0.201&0.550& & & 0.637 & 0.191 & 0.888 & 0.252\\
\midrule
\multirowbt{1}{*}{1}&0.341&0.530& & & 0.661 & 0.187 & 0.860 & 0.243\\
\midrule
\multirowbt{1}{*}{2}&0.525&0.513& 0.390& 0.322& 0.672 & 0.177 & 0.851 & 0.243  \\
\midrule
\multirowbt{1}{*}{5}&0.760&0.532& 0.570& 0.236& 0.671 & 0.151 & 0.850 & 0.254\\
\midrule
\multirowbt{1}{*}{10}&0.864&0.548& 0.679& 0.184&0.661 & 0.119 & 0.854 & 0.270\\
\midrule
\multirowbt{1}{*}{20}&0.919&0.554& & &0.645 & 0.079 & 0.859 & 0.290\\
\bottomrule\end{tabular}
\label{31t}
\end{table*}

\subsection{$4/1$ resonance}
\begin{figure*}[!htp]
\begin{center}
\includegraphics[width=10cm,height=10cm]{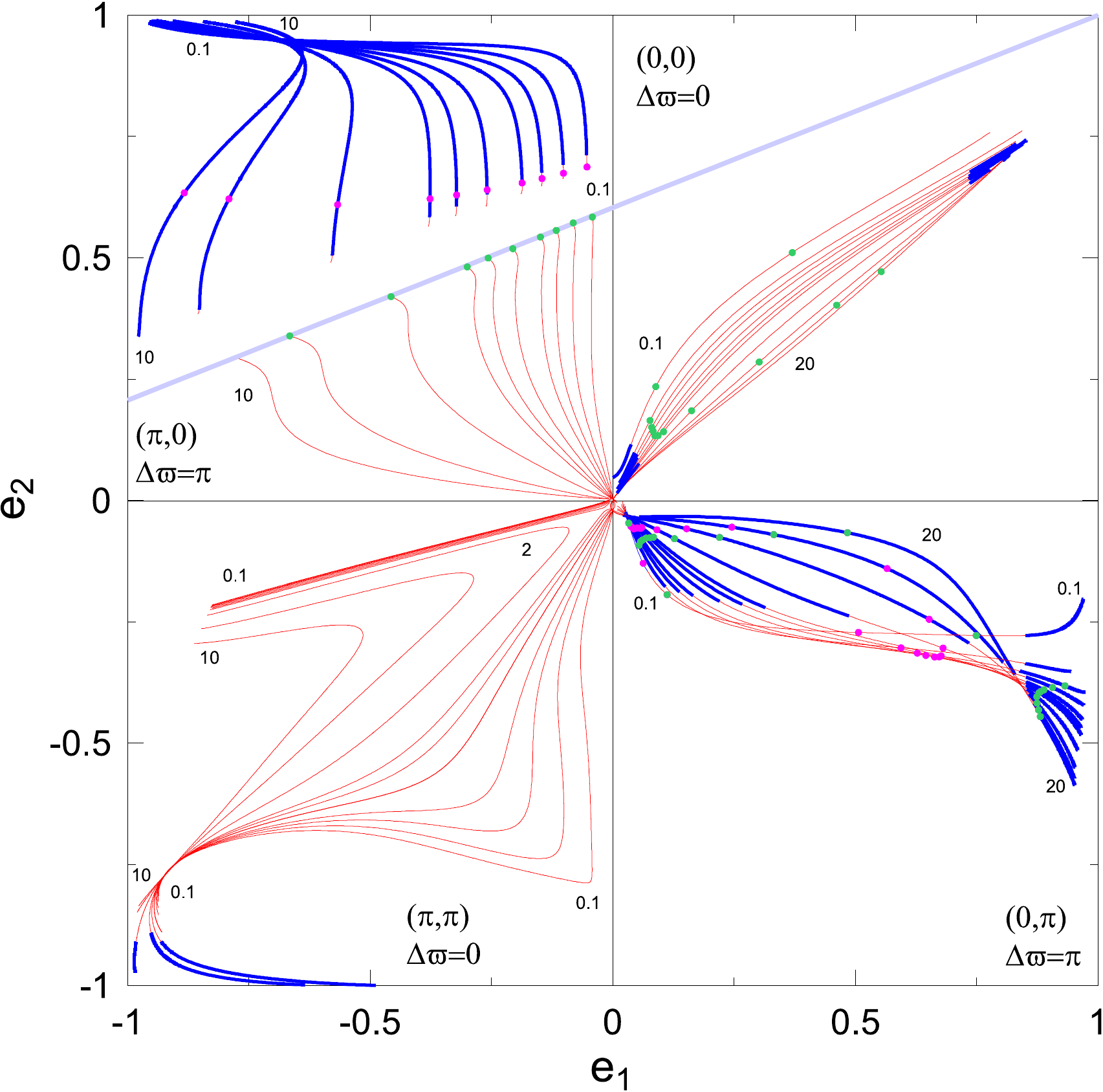} 
\end{center}
\caption{Presentation of planar families of symmetric periodic orbits in $4/1$ resonance, as in Figure \ref{43}.}
\label{41}
\end{figure*}

Extrasolar planetary systems, which are trapped in $4/1$ resonance, are e.g. HD $102272$ \citep{nied09}, HD $108874$ \citep{gozd06,lib07} and GJ $667$C \citep{angl12}. In Figure \ref{41}, we depict the families, indicating the stability and the v.c.o. at each configuration.

Similarly to the 3/1 resonance, every configuration seems to have families that bifurcate from circular periodic orbits. Segments with stable orbits are apparent in every configuration, but only for the cases $(0,0)$ and $(0,\pi)$ we observe stability at nearly circular orbits. There are stable v.c.o. in configurations $(0,\pi)$ and $(\pi,0)$ and they are given in Table \ref{41t}. In configuration (0,0) all v.c.o are unstable, while in $(\pi,\pi)$ they do not exist. 

\begin{table*}[!htp]
\caption{Eccentricity values of v.c.o. in $4/1$ resonance.}

\begin{tabular}[b]{lllllllll}
\toprule
 & $\widehat{F}^{4/1}_{(0,\pi)}$ & &$\widehat{G}^{4/1}_{(0,\pi)}$ & & $\widehat{F}^{4/1}_{(\pi,0)}$ & & $\widehat{G}^{4/1}_{(\pi,0)}$& \\
 \cmidrule{2-9}
$\rho$ & $e_1$ & $e_2$ & $e_1$ & $e_2$& $e_1$ & $e_2$ & $e_1$ & $e_2$\\
\midrule
\multirowbt{3}{*}{0.1} & 0.032 & 0.046 & 0.062 & 0.129 &0.053&0.686& 0.042&0.583\\
\cmidrule{2-9}
 & 0.506 & 0.272 & 0.111 & 0.194 &&&&\\
\cmidrule{2-5}
 &  &  & 0.748 & 0.278 &&&&\\
\midrule
\multirowbt{2}{*}{0.2} & 0.037 & 0.054 & 0.054 & 0.092&0.101&0.674&0.081& 0.571 \\
\cmidrule{2-9}
 & 0.593 & 0.304 &  &  &&&&\\
\midrule
\multirowbt{2}{*}{0.3} & 0.042 & 0.058 & 0.057 & 0.085&0.146&0.663&0.116 &0.556  \\
\cmidrule{2-9}
 & 0.627 & 0.314 & 0.931 & 0.382&&&&\\
\midrule
\multirowbt{2}{*}{0.4} & 0.045 & 0.057 & 0.061 & 0.081&0.186&0.653& 0.149& 0.542 \\
\cmidrule{2-9}
 & 0.644& 0.319 & 0.906 & 0.385&&&&\\
\midrule
\multirowbt{2}{*}{0.6} & 0.050 & 0.056 & 0.068 & 0.077&0.259&0.639&0.206 &0.518 \\
\cmidrule{2-9}
 & 0.663 & 0.322 & 0.888 & 0.390&&&&\\
\midrule
\multirowbt{2}{*}{0.8} & 0.056 & 0.058 & 0.076 & 0.076&0.321&0.629& 0.257& 0.499 \\
\cmidrule{2-9}
 & 0.672 & 0.322 & 0.881 & 0.393 &&&&\\
\midrule
\multirowbt{2}{*}{1} & 0.059 & 0.056 & 0.083 & 0.076&0.376&0.621&0.300 &0.480  \\
\cmidrule{2-9}
 & 0.676 & 0.320 & 0.878 & 0.396 &&&&\\
\midrule
\multirowbt{2}{*}{2} & 0.090 & 0.060 & 0.126 & 0.078&0.567&0.609& 0.457& 0.419 \\
\cmidrule{2-9}
 & 0.680 & 0.304 & 0.873 & 0.404 &&&&\\
\midrule
\multirowbt{2}{*}{5} & 0.151 & 0.058 & 0.219 & 0.076 &0.791&0.621& 0.665& 0.338\\
\cmidrule{2-9}
 & 0.651 & 0.245 & 0.874 & 0.418 &&&&\\
\midrule
\multirowbt{2}{*}{10} & 0.245 & 0.055 & 0.331 & 0.070 &0.883&0.633& & \\
\cmidrule{2-7}
 & 0.565 & 0.140 & 0.877 & 0.431 &&&&\\
\midrule
\multirowbt{2}{*}{20} &  &  & 0.483 & 0.066 &&&&\\
\cmidrule{4-5}
 &  &  & 0.881 & 0.445 &&&&\\
\bottomrule\end{tabular}
\label{41t}
\end{table*}

\section{Spatial families of periodic orbits}

We herein present some indicative spatial families of periodic orbits, which bifurcate from stable v.c.o.. Such spatial families can start as stable or unstable. But families which bifurcate from an unstable v.c.o. start always as unstable. Of course, the stability type can change along the family. Along the continuation in the third dimension the symmetry does not change, but their starting configuration $(\theta_1,\theta_2)$ of the planar v.c.o. may change along the spatial family. We remark that a family is named after the configuration of the v.c.o. it started from and it is presented in the projection space $(e_1,e_2,\Delta i)$ (see Sect. \ref{SecPres}). 

\subsection{$4/3$ resonance}
\label{S43}
\begin{figure*}[!htp]
\begin{center}
$\begin{array}{ccc}
\includegraphics[width=6cm,height=6cm]{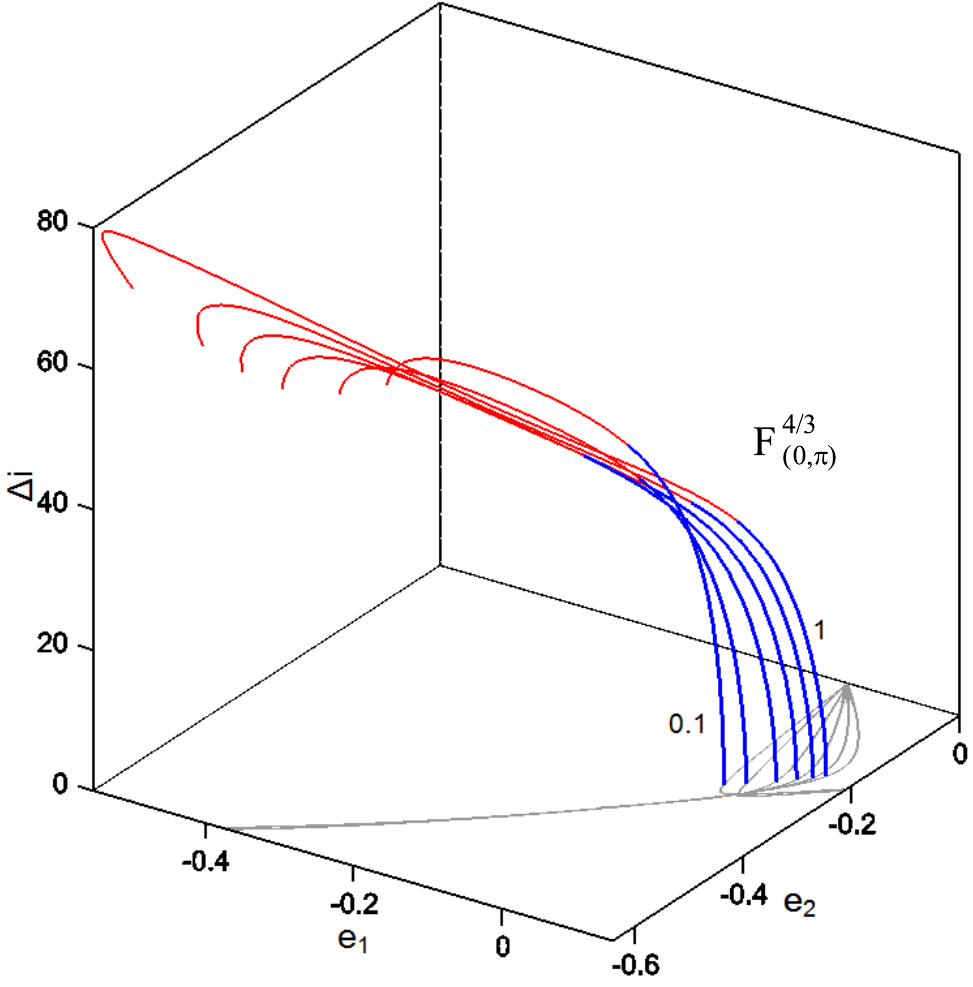}  & \quad&
\includegraphics[width=6cm,height=6cm]{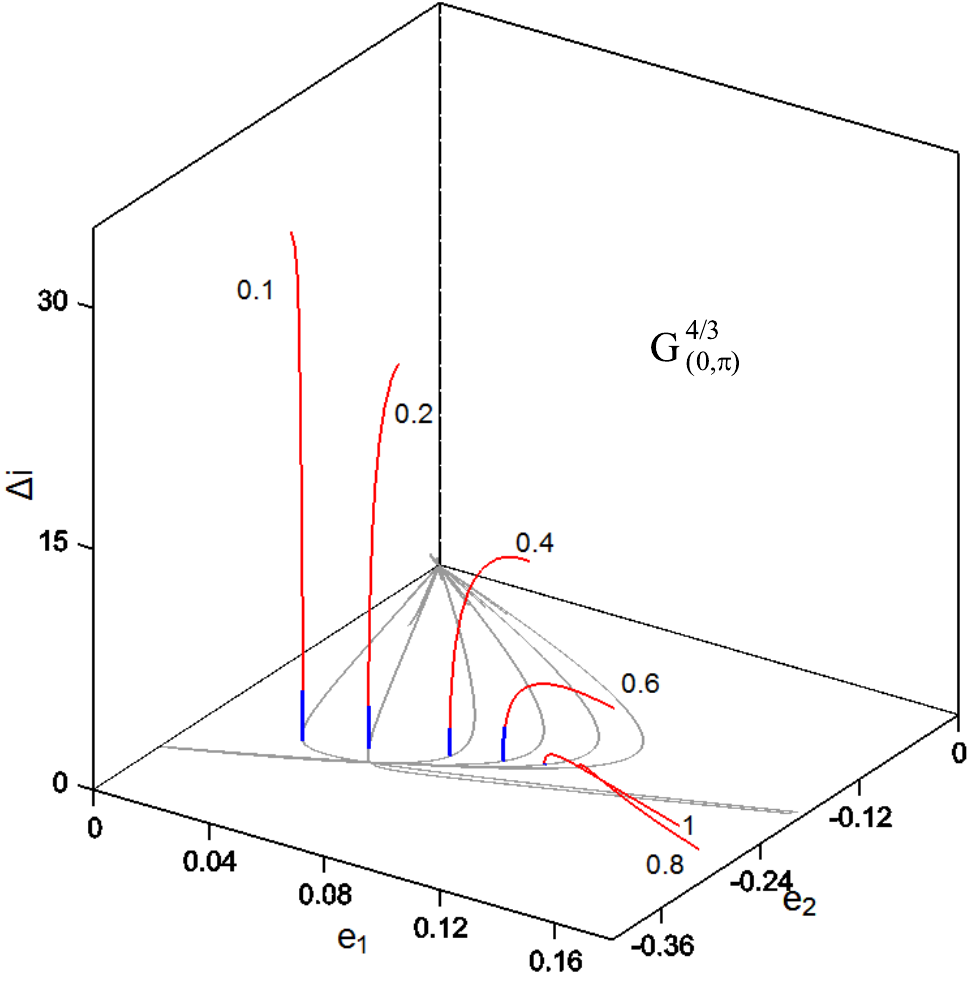} \\
\textnormal{(a)} & \quad & \textnormal{(b)} \\
\includegraphics[width=6cm,height=6cm]{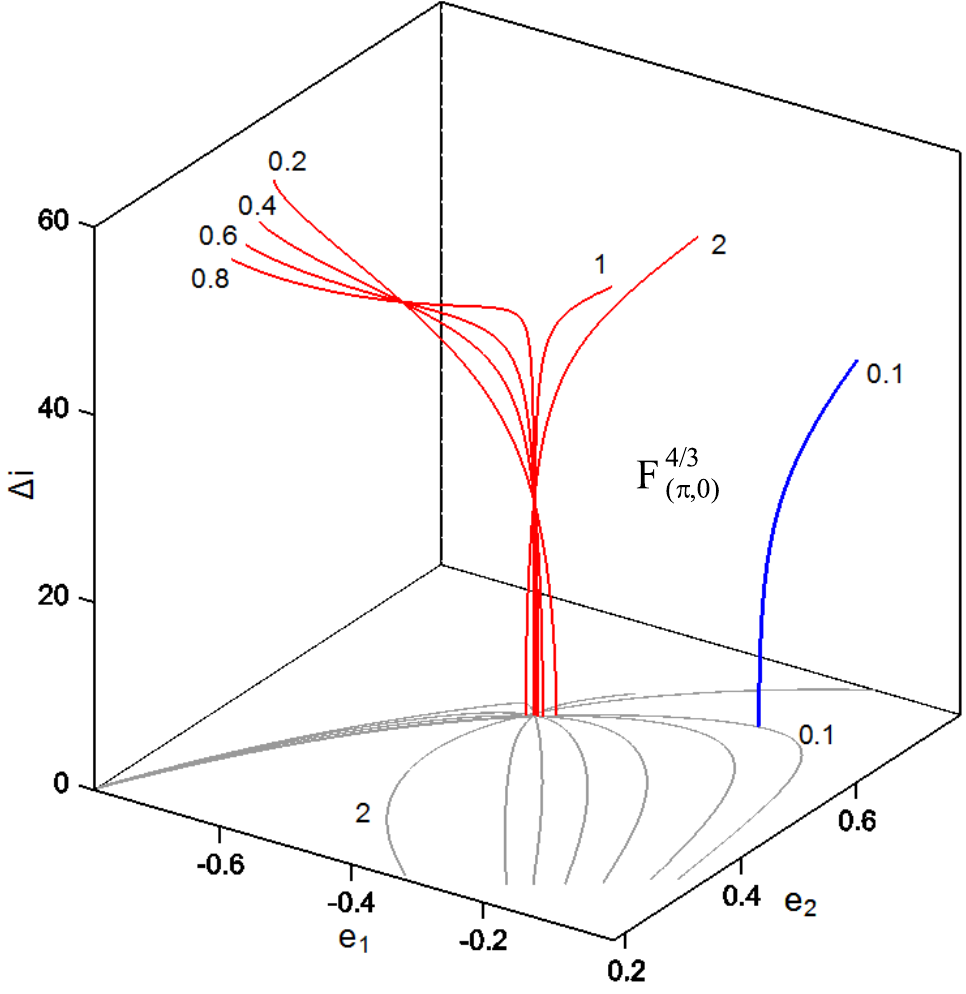}  & \quad&
\includegraphics[width=6cm,height=6cm]{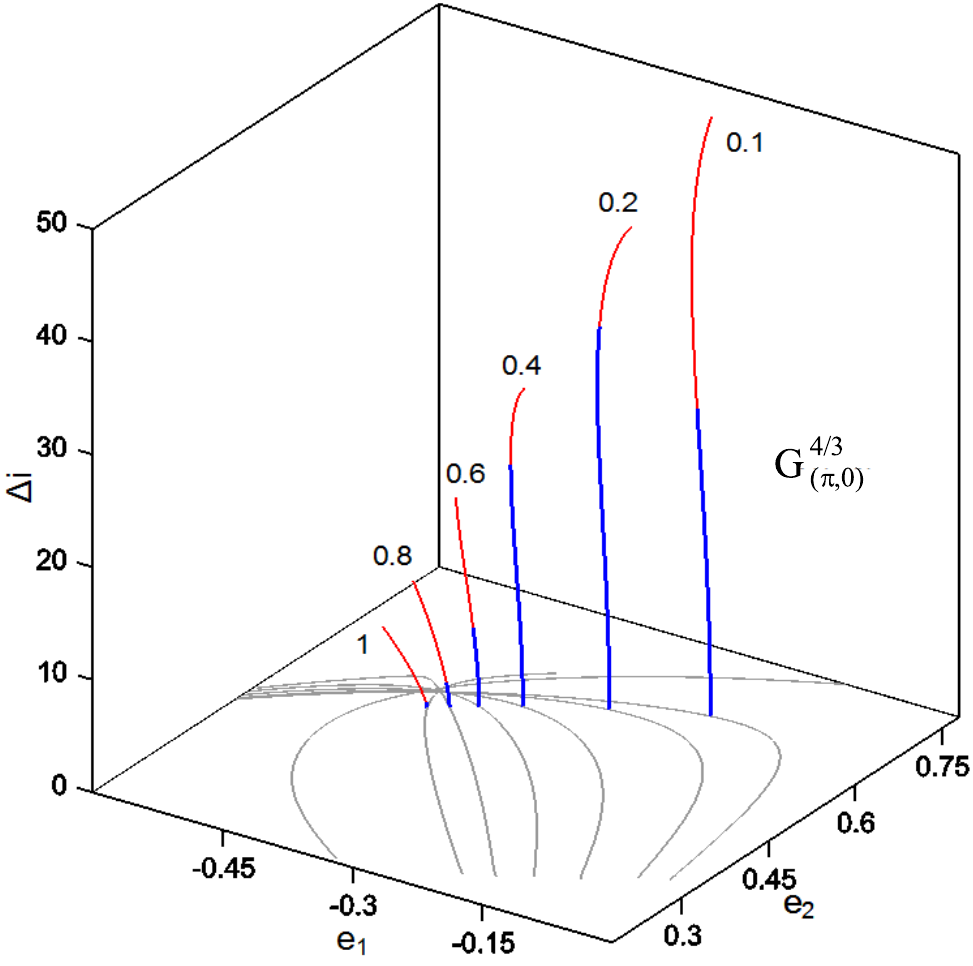} \\
\textnormal{(c)} & \quad & \textnormal{(d)} \\
\end{array} $
\end{center}
\caption{Spatial families of symmetric periodic orbits in $4/3$ resonance for the indicated mass ratios. Both configurations $(0,\pi)$ and $(\pi,0)$ have $xz$-symmetric (panels {\bf a,c}) and $x$-symmetric (panels {\bf b,d}) families of periodic orbits. Blue and red coloured lines stand for stable and unstable orbits, respectively. The planar families, where they bifurcate from, are also presented with gray colour.}
\label{43s1}
\end{figure*}

In Figure \ref{43s1}, we present families of $3D$ periodic orbits bifurcating from stable v.c.o. of the particular resonance. The configuration of these v.c.o. is $(0,\pi)$ or $(\pi,0)$ and we obtain spatial families $F$ and $G$ (symmetries $xz$ and $x$, respectively).
We observe that in the projection space the families start with an increasing mutual inclination $\Delta i$, while the eccentricities do not vary significantly. This feature appears in most cases. For larger values of $\Delta i$, eccentricities may also vary significantly along the families.     

\begin{figure}[H]
\begin{center}
\includegraphics[width=7cm,height=7cm]{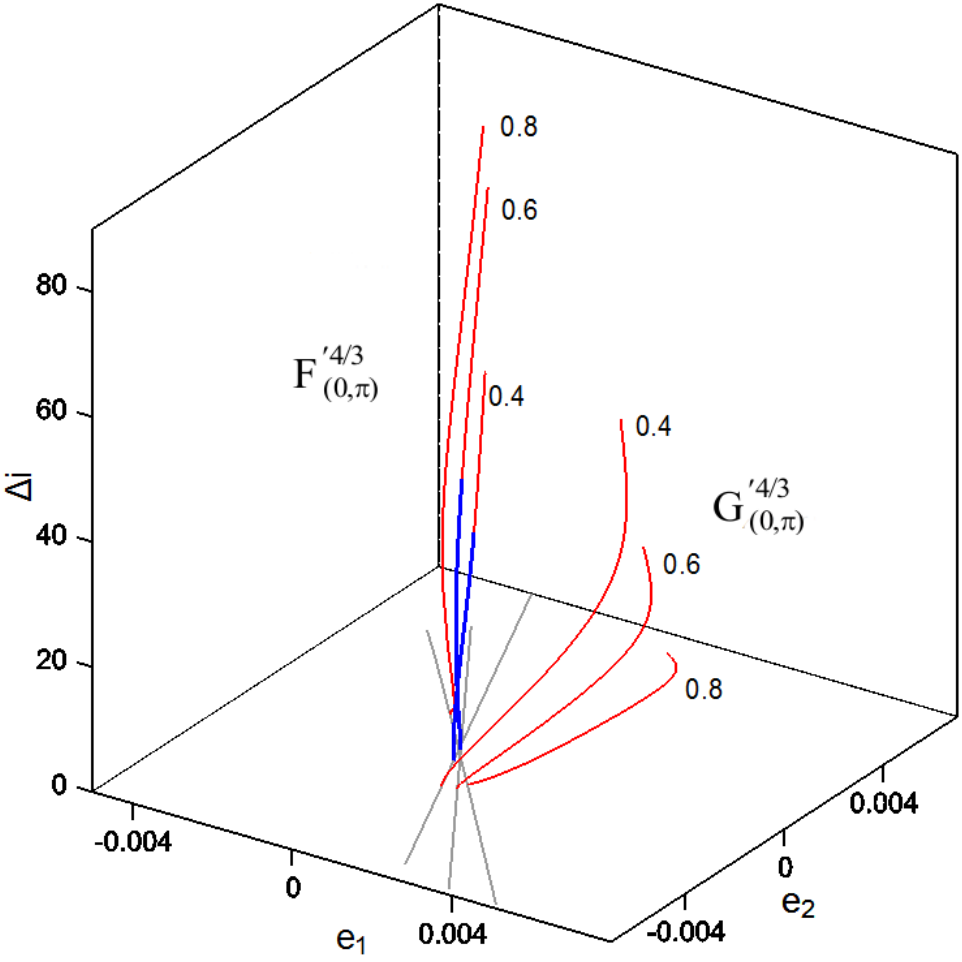}  
\end{center}
\caption{Spatial families of periodic that bifurcate from circular v.c.o. close to the 4/3 resonance.}
\label{43c}
\end{figure}

In Figure \ref{43c}, we depict the spatial families of periodic orbits that are continued from the circular v.c.o. (see Sect. \ref{P43}) and we denote them with a prime. The ${G}^{\prime4/3}_{(0,\pi)}$ families are unstable, whilst the ${F}^{\prime4/3}_{(0,\pi)}$ families are stable up to a mutual inclination of $28^\circ$ and $35^\circ$ for mass ratios $\rho=0.4$ and $\rho=0.6$, respectively. But, for $\rho=0.8$, ${F}^{\prime4/3}_{(0,\pi)}$ becomes totally unstable. Along the stable segments the eccentricities do not vary significantly and the planetary orbits remain almost circular. This is the only case of circular inclined periodic orbits we found. 

\begin{figure}[H]
\begin{center}
\includegraphics[width=6cm,height=6cm]{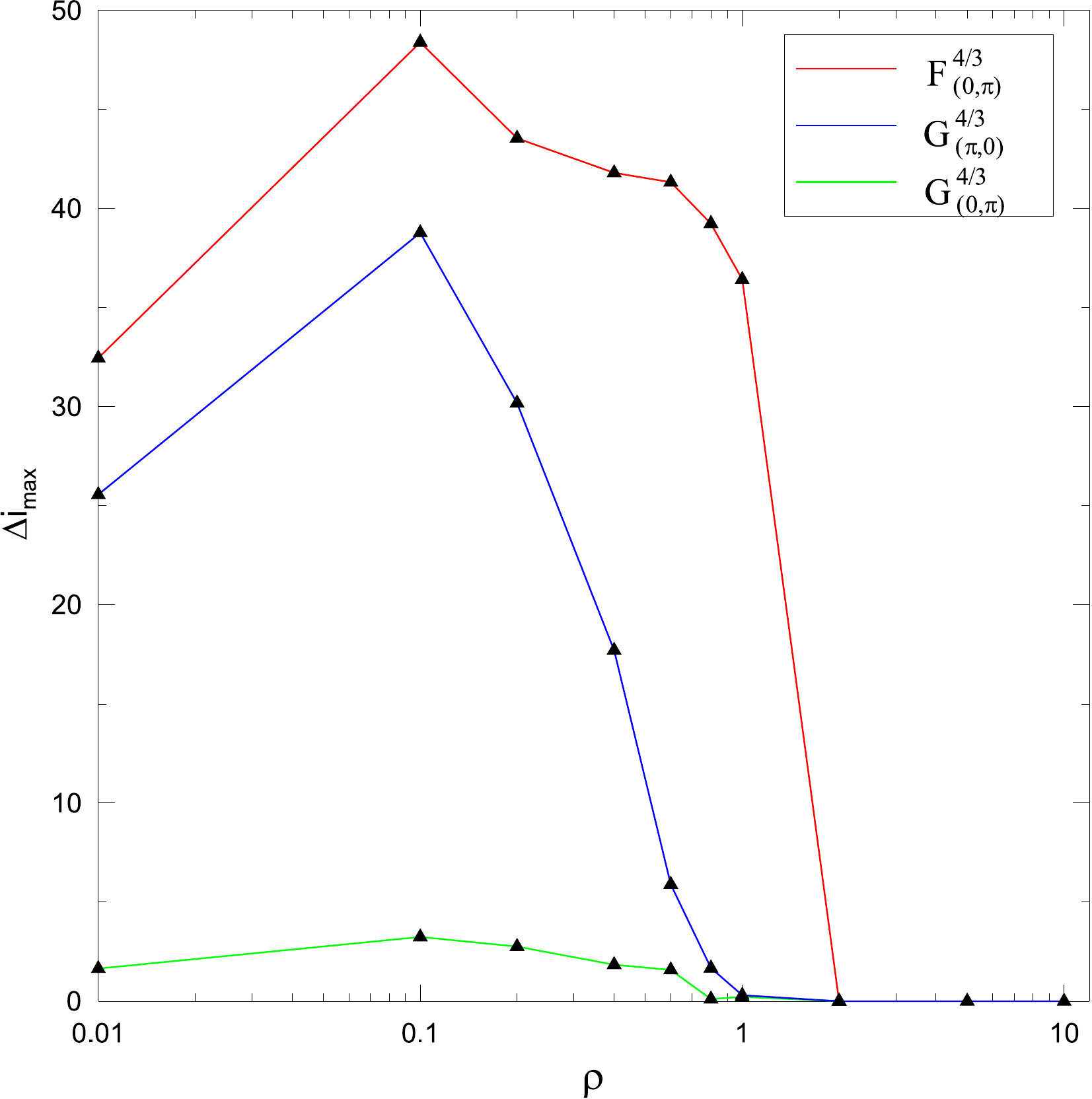}  
\end{center}
\caption{Maximum values of mutual inclination, $\Delta i_{max}$, up to which we obtain stable periodic orbits along the indicated families as a function of the mass ratio $\rho$.}
\label{43di}
\end{figure}

In Figure \ref{43di}, we show the maximum value of mutual inclination up to which stable orbits exist as a function of the mass ratio, $\rho$. We consider the cases where the families start having stable regions. We observe that the maximum mutual inclination of all the symmetries has the same behaviour, namely it reaches a maximum value ($48^{\circ}$, $5^{\circ}$, $38^{\circ}$, corresponding to Figs. \ref{43s1}{\bf a},{\bf b},{\bf d}, accordingly) and then, it decreases monotonically down to zero.

\subsection{$3/2$ resonance}
\begin{figure*}[!htp]
\begin{center}
$\begin{array}{ccc}
\includegraphics[width=6cm,height=6cm]{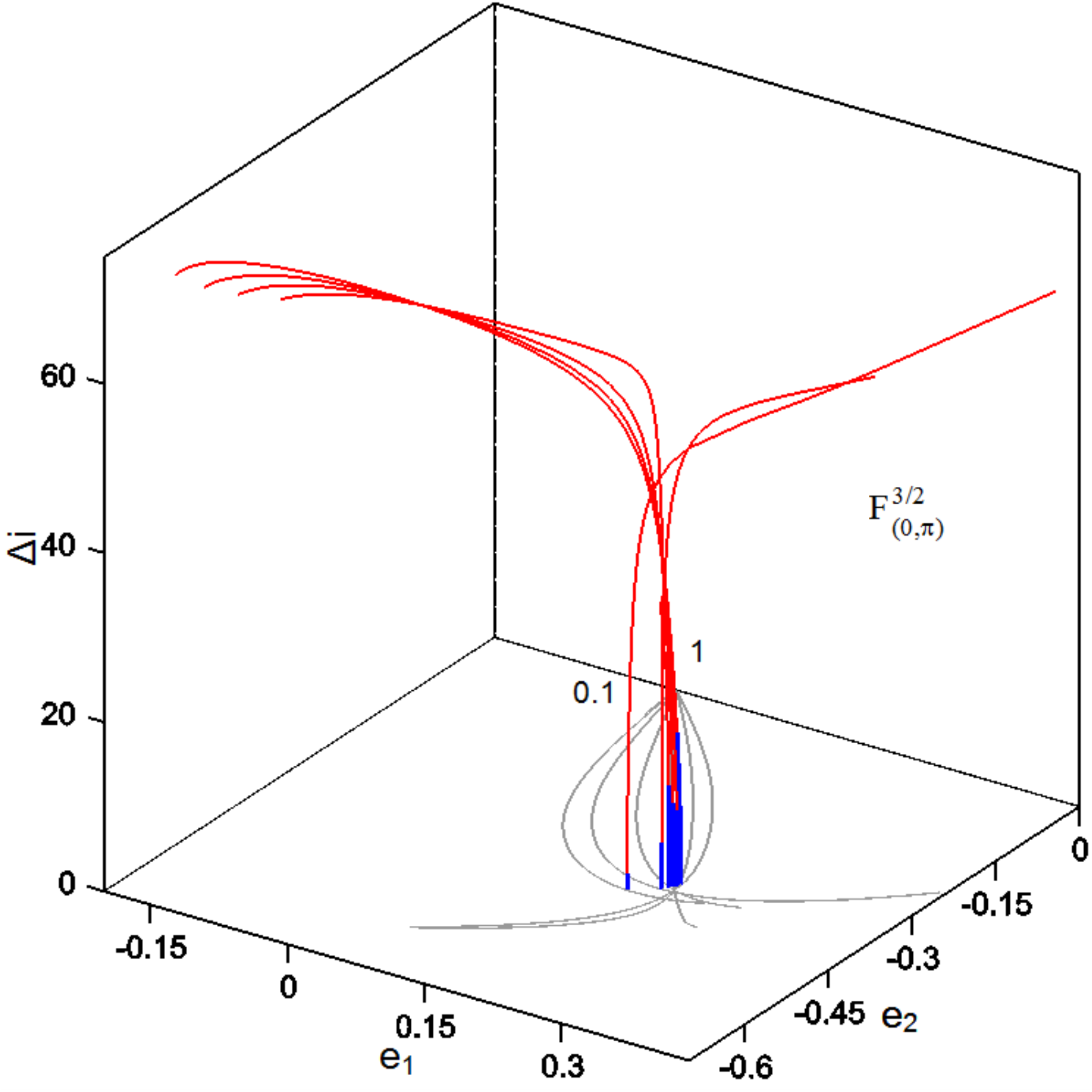}  & \quad&
\includegraphics[width=6cm,height=6cm]{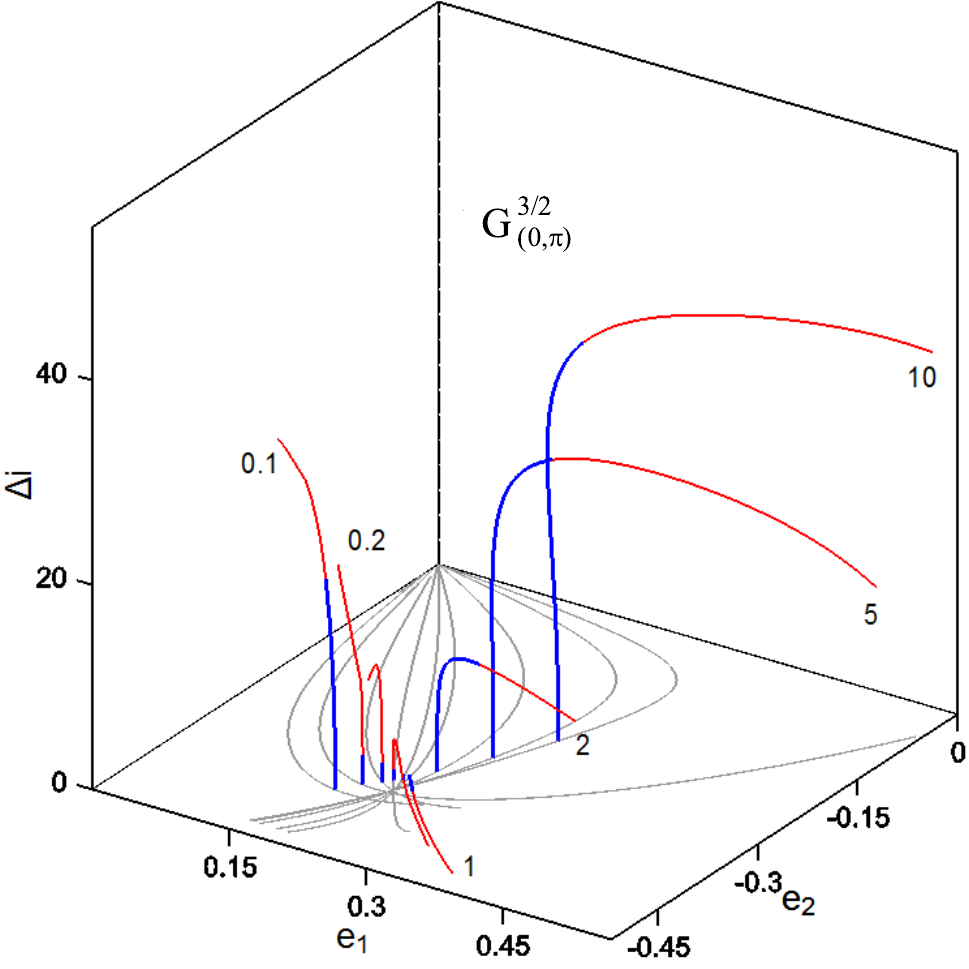} \\
\textnormal{(a)} & \quad & \textnormal{(b)} \\
\includegraphics[width=6cm,height=6cm]{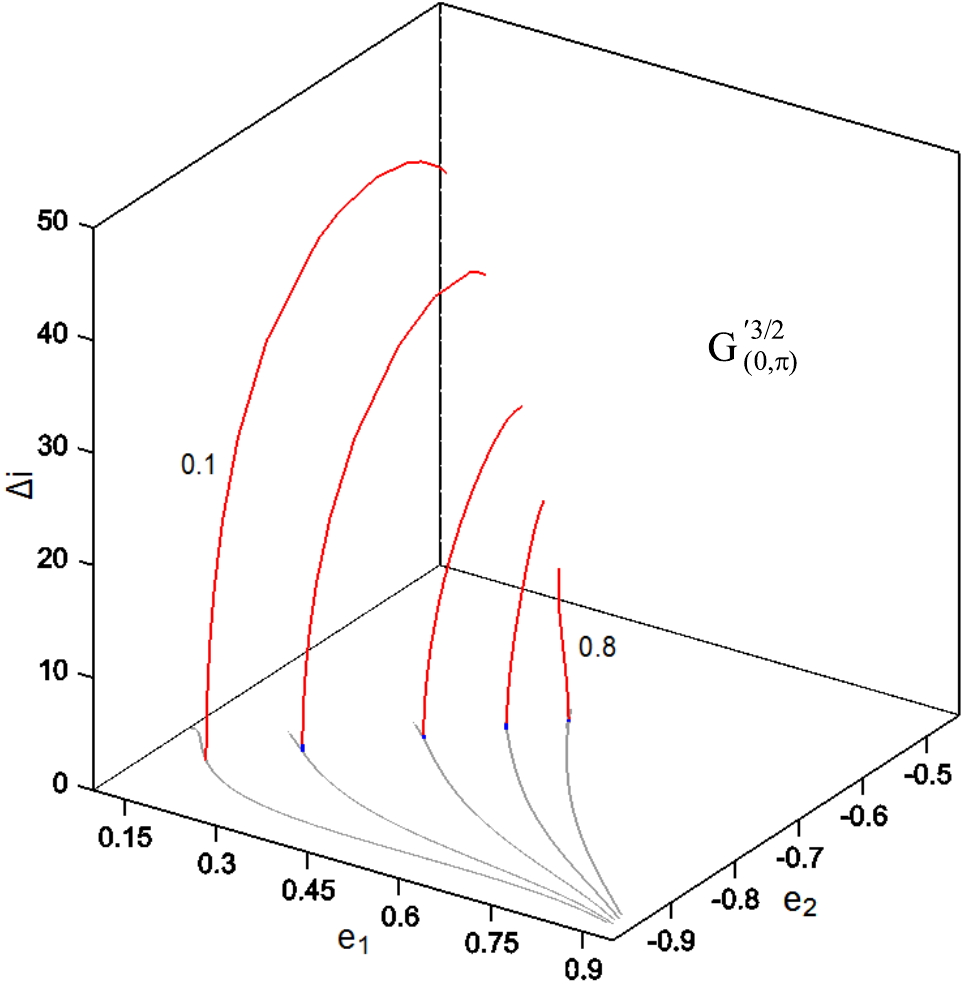}  & \quad&
\includegraphics[width=6cm,height=6cm]{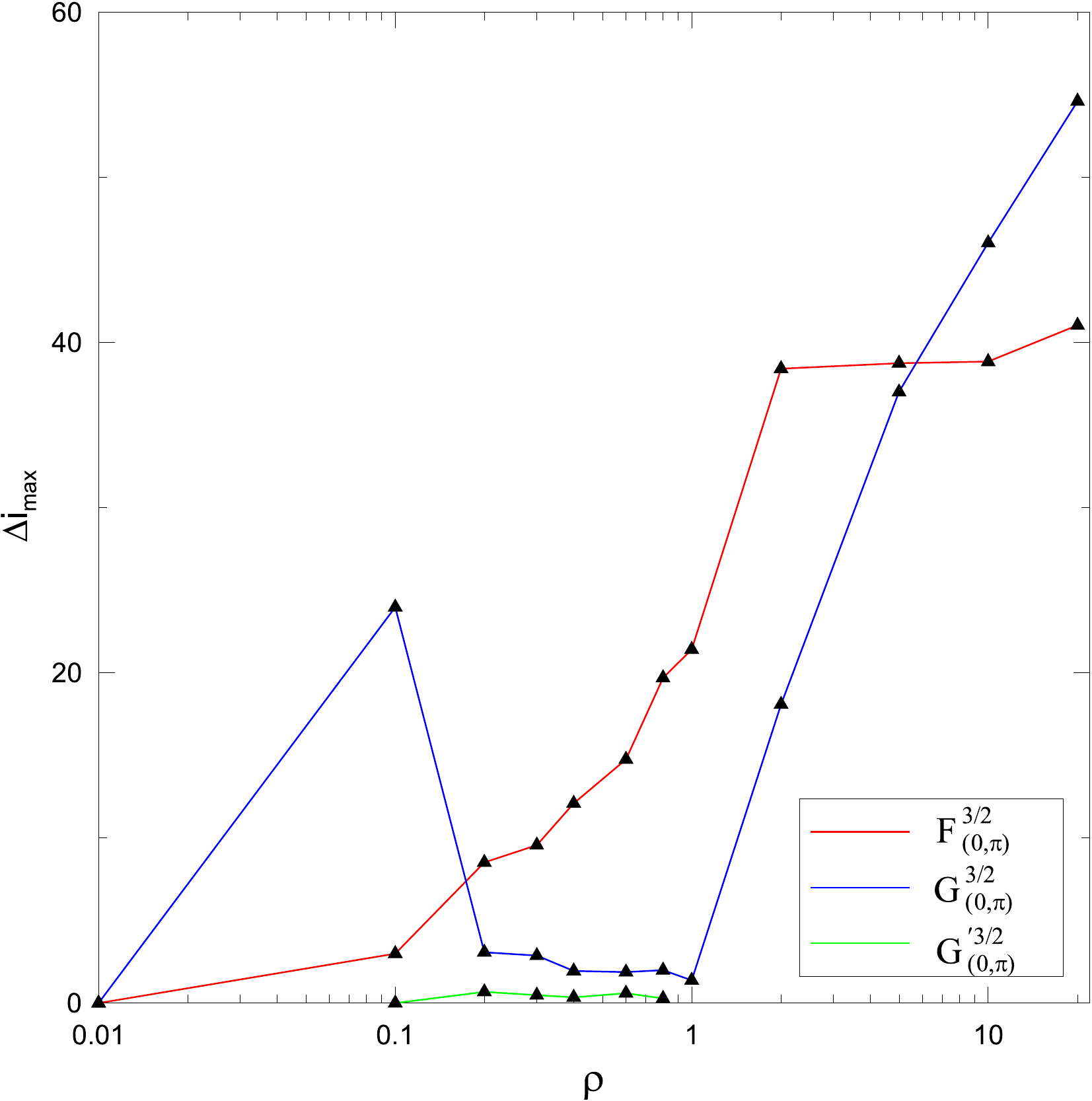} \\
\textnormal{(c)} & \quad & \textnormal{(d)} 
\end{array} $
\end{center}
\caption{Spatial families of symmetric periodic orbits in $3/2$ resonance for various mass ratios. {\bf a} ${F}^{3/2}_{(0,\pi)}$, {\bf b} ${G}^{3/2}_{(0,\pi)}$ and {\bf c} ${G}^{\prime3/2}_{(0,\pi)}$ families of periodic orbits. {\bf d} Maximum values of mutual inclination reached along the families as a function of the mass ratio $\rho$.}
\label{32s1}
\end{figure*}

In Fig. \ref{32s1}, we present the spatial groups of families of periodic orbits that belong to the configuration $(0,\pi)$. In panels {\bf a} and {\bf b}, the ${F}^{3/2}_{(0,\pi)}$ and ${G}^{3/2}_{(0,\pi)}$ families of periodic orbits are depicted. In panel {\bf c}, we present a second group, ${G}^{\prime3/2}_{(0,\pi)}$, of $x$-symmetry families emerging from the group of planar families "above" the collision line (see Fig. \ref{32}). All of them start having stable regions, but the stability does not exceed the value $\Delta i \approx 1^\circ$. In panel {\bf d}, we observe that the stability of families ${F}^{3/2}_{(0,\pi)}$ appears up to a mutual inclination value $\Delta i_{max}$, which increases as $\rho$ increases and reaches the value of $41^\circ$ for $\rho=20$. The families ${G}^{3/2}_{(0,\pi)}$ show stability up to large mutual inclination values for $\rho>1$ and for $\rho=0.1$ where $\Delta i_{max}=24^\circ$. As we mentioned above, the families ${G}^{\prime3/2}_{(0,\pi)}$ are, practically, totally unstable.   

\subsection{$5/2$ resonance}
\begin{figure*}[!htp]
\begin{center}
$\begin{array}{ccc}
\includegraphics[width=6cm,height=6cm]{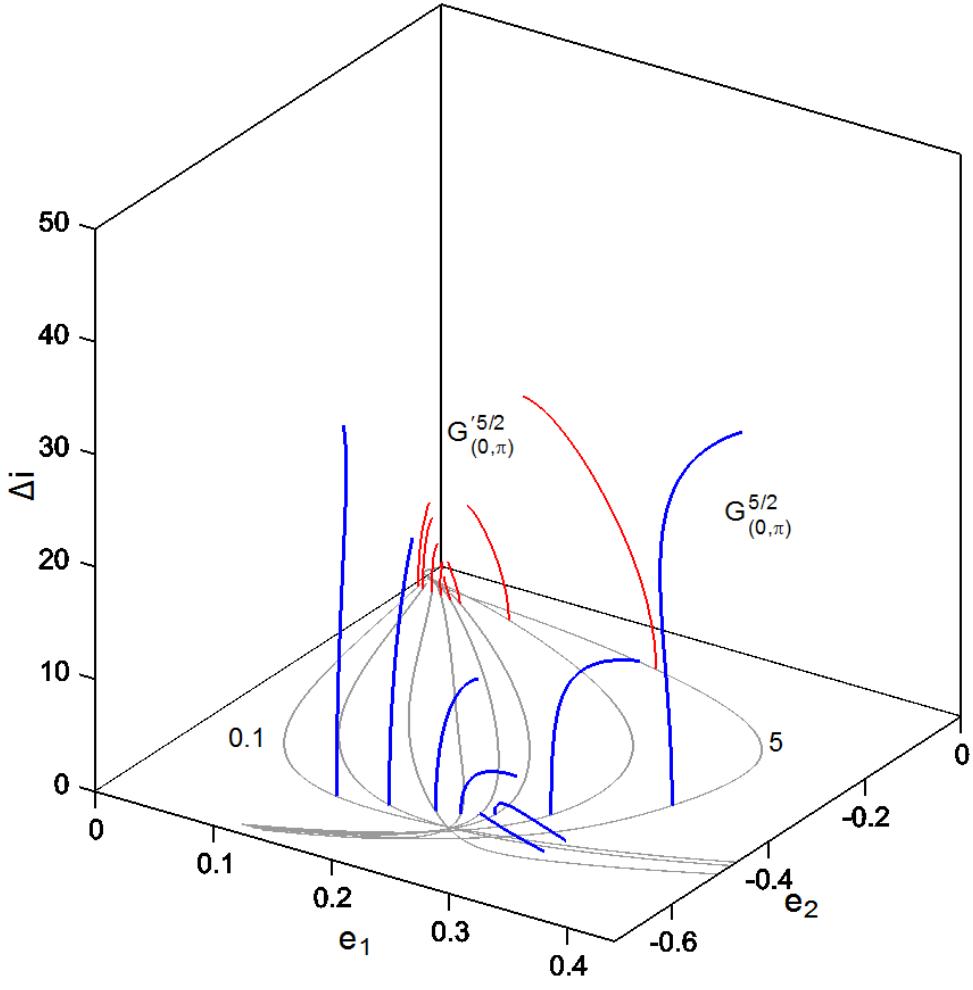}  & \quad&
\includegraphics[width=6cm,height=6cm]{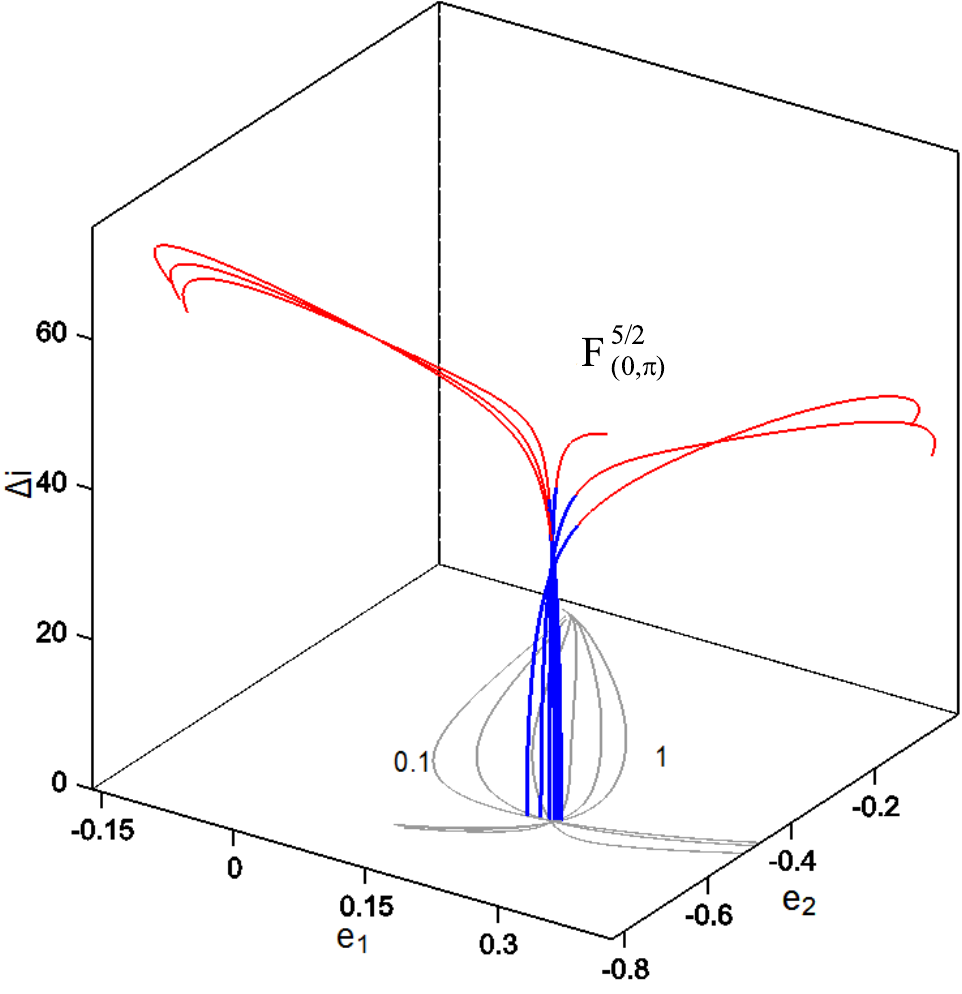} \\
\textnormal{(a)} & \quad & \textnormal{(b)}\\ 
\includegraphics[width=6cm,height=6cm]{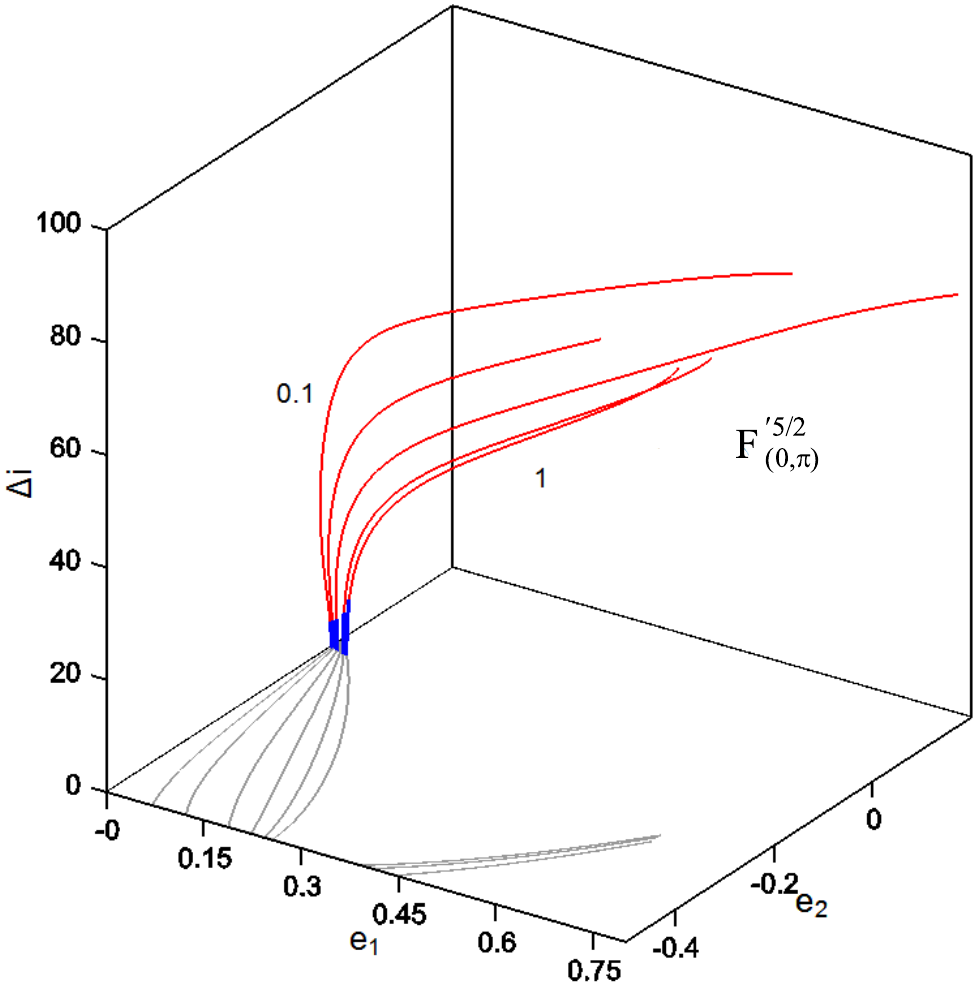}  & \quad&
\includegraphics[width=6cm,height=6cm]{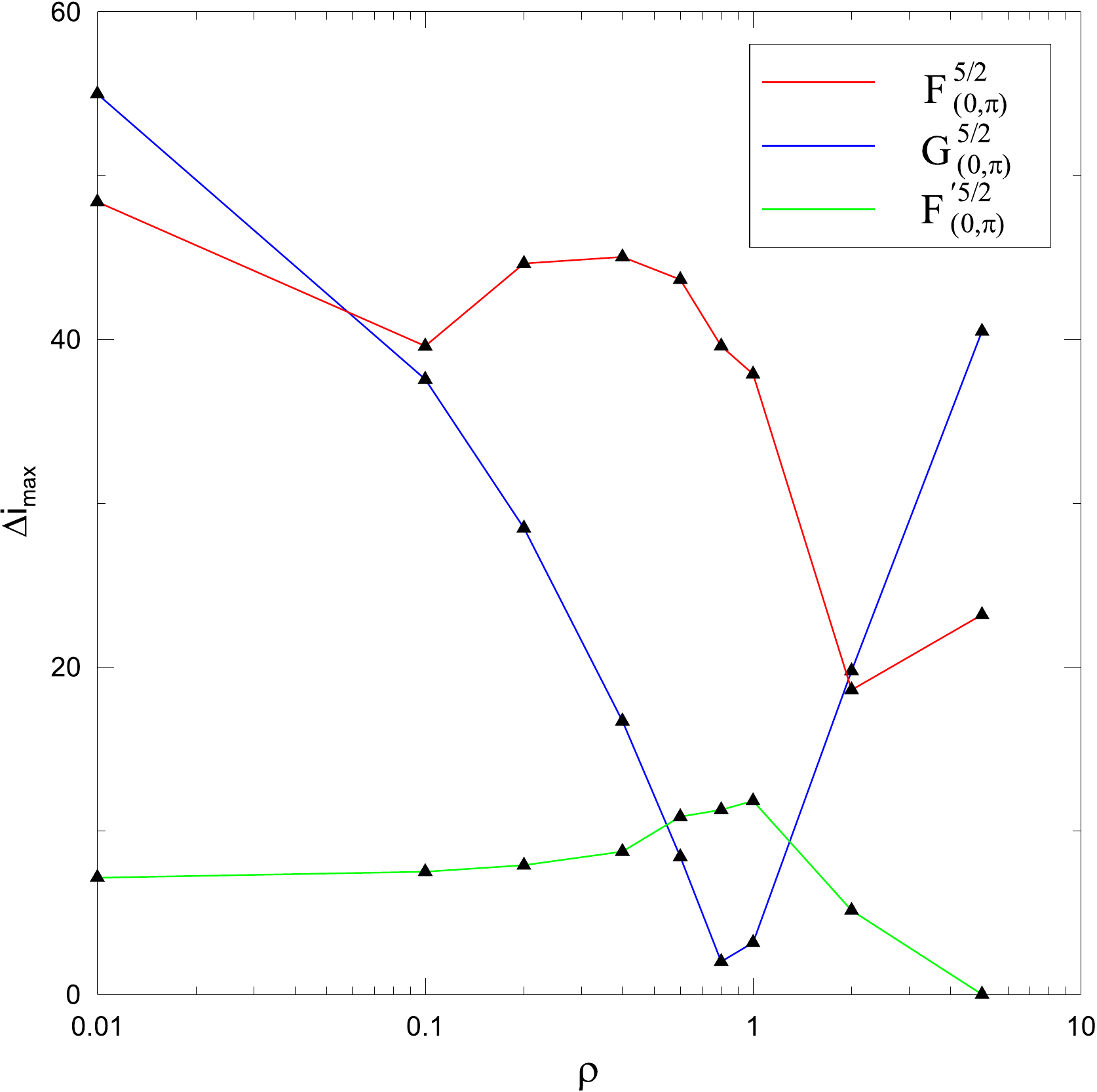} \\
\textnormal{(c)} & \quad & \textnormal{(d)} 
\end{array} $
\end{center}
\caption{{\bf a} Families ${G}^{5/2}_{(0,\pi)}$ {\bf b} Families ${F}^{5/2}_{(0,\pi)}$  {\bf c} Families ${F}^{\prime 5/2}_{(0,\pi)}$  {\bf d} Maximum values of mutual inclination reached along the families as a function of the mass ratio $\rho$.}
\label{52b}
\end{figure*}

In Fig. \ref{52b}{\bf a}, we present the ${G}^{5/2}_{(0,\pi)}$ and ${G}^{\prime 5/2}_{(0,\pi)}$ families of periodic orbits generated from the two pairs of v.c.o. of the planar families (see Sect. \ref{P52}). We observe that ${G}^{5/2}_{(0,\pi)}$ consist of stable periodic orbits, while ${G}^{\prime 5/2}_{(0,\pi)}$ are totally unstable. The maximum value of mutual inclination that the planets of the stable orbits reach is of $55^\circ$ (see Fig. \ref{52b}{\bf d}, blue line). However, for $\rho\approx 1$, the stability is evident only for very small mutual inclination values.

In panels {\bf b} and {\bf c} of Fig. \ref{52b}, we show the families ${F}^{5/2}_{(0,\pi)}$ and ${F}^{\prime 5/2}_{(0,\pi)}$, respectively. For $\rho<2$ families ${F}^{5/2}_{(0,\pi)}$ are stable up to large inclination values (up to $45^\circ$ for $\rho=0.4$). In their stable segments the eccentricities are almost constant. Families ${F}^{\prime 5/2}_{(0,\pi)}$ start as stable, but their stability does not exceed $12^\circ$ (for $\rho=1$). The maximum mutual inclination value for each mass ratio is shown in panel {\bf d}.

\subsection{$3/1$ resonance}
\label{331}
\begin{figure}[H]
\begin{center}
$\begin{array}{ccc}
\includegraphics[width=6cm,height=6cm]{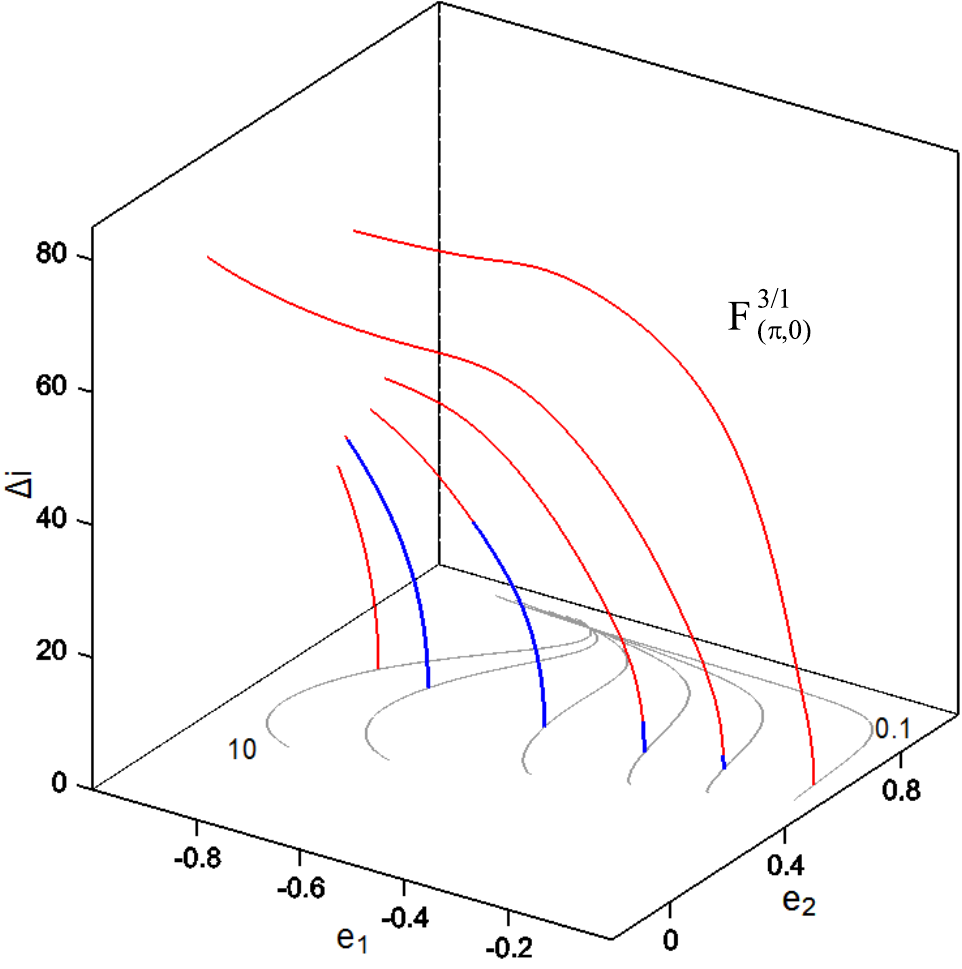} &\quad&
\includegraphics[width=6cm]{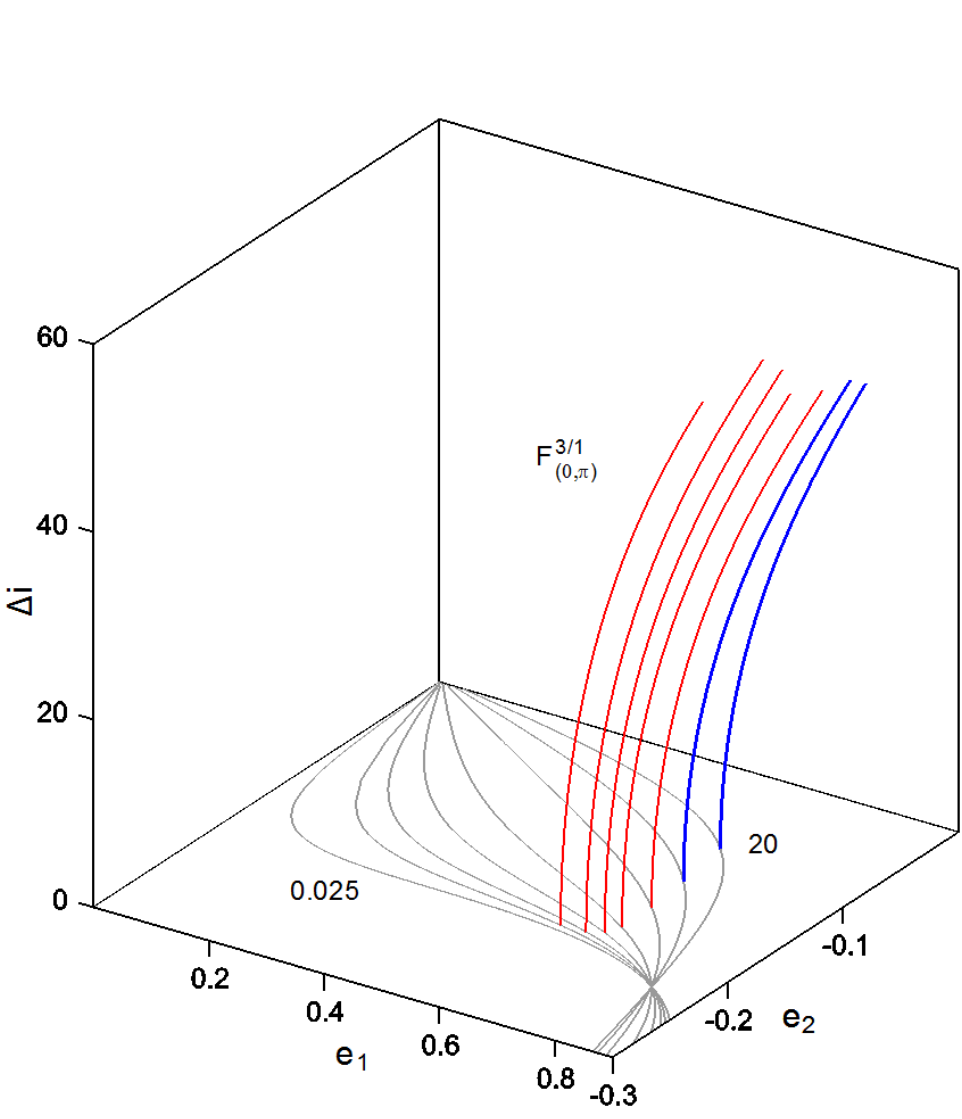} \\
\textnormal{(a)}&\quad&\textnormal{(b)} 
\end{array} $
\end{center}
\caption{{\bf a} Families ${F}^{3/1}_{(\pi,0)}$ and {\bf b} ${F}^{3/1}_{(0,\pi)}$.}
\label{31s2}
\end{figure}

\begin{figure}[H]
\begin{center}
\includegraphics[width=6cm,height=6cm]{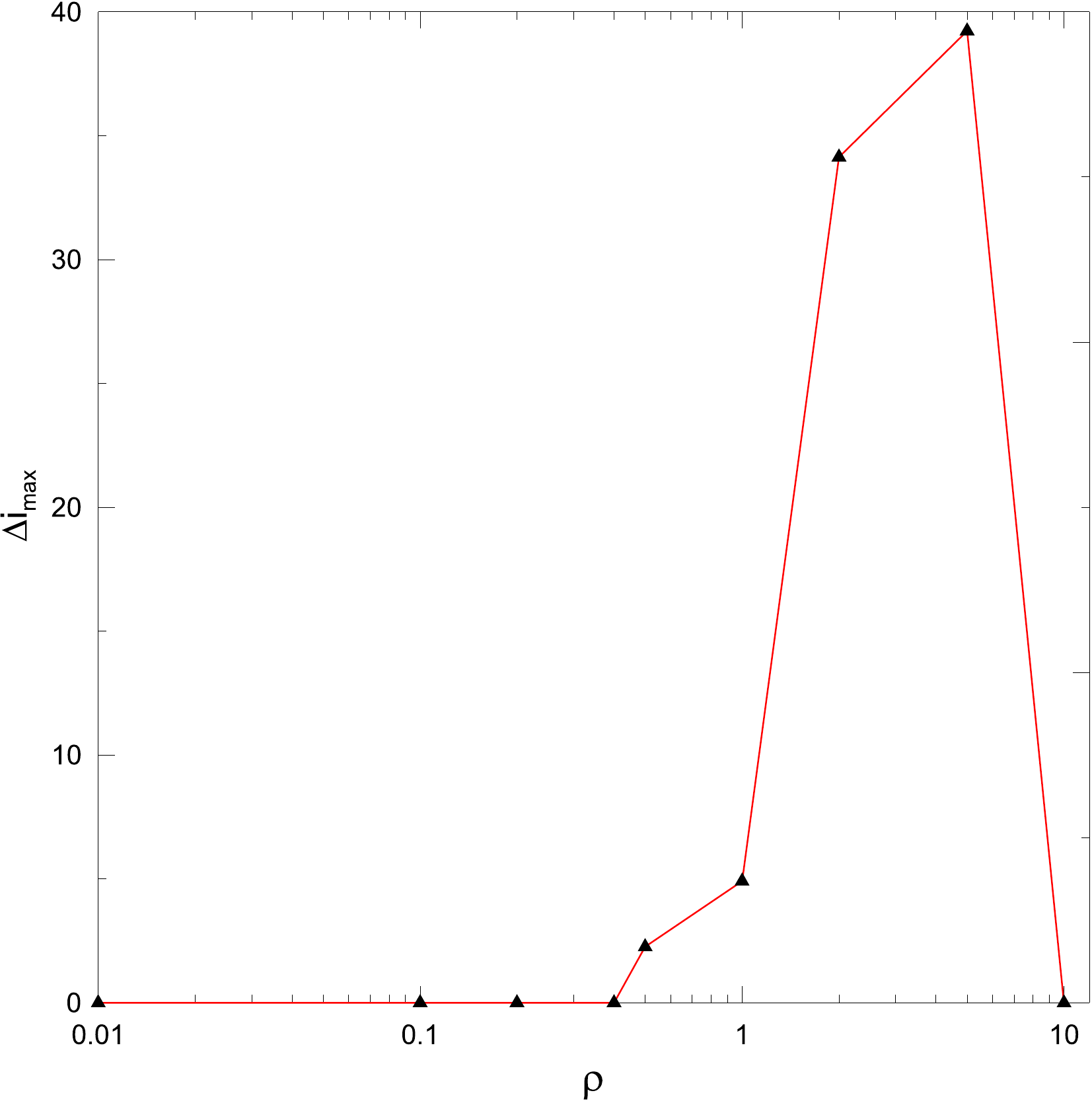}
\end{center}
\caption{Maximum values of mutual inclination along the family ${F}^{3/1}_{(\pi,0)}$ as a function of the mass ratio $\rho$.}
\label{di31}
\end{figure}

In Fig. \ref{31s2}, we present groups of the spatial families ${F}^{3/1}_{(\pi,0)}$  and ${F}^{3/1}_{(0,\pi)}$ (panels {\bf a} and {\bf b}, respectively), for various mass ratios. In Fig. \ref{di31}, we show the $\Delta i_{max}$ which is reached by the orbits of the families ${F}^{3/1}_{(\pi,0)}$ as the planetary mass ratio $\rho$ varies. For, approximately, $\rho<0.5$ and $\rho\geq 10$ the families have no stable orbits. The highest value of $\Delta i_{max}$ is obtained for $\rho=5$ and is equal to $39^\circ$. For the families ${F}^{3/1}_{(0,\pi)}$ we found that they are totally unstable for $\rho\leq5$,  but for $\rho\geq10$ we obtain stability up to mutual inclination of approximately $50^\circ$.

\begin{figure}[H]
\begin{center}
\includegraphics[width=7cm,height=7cm]{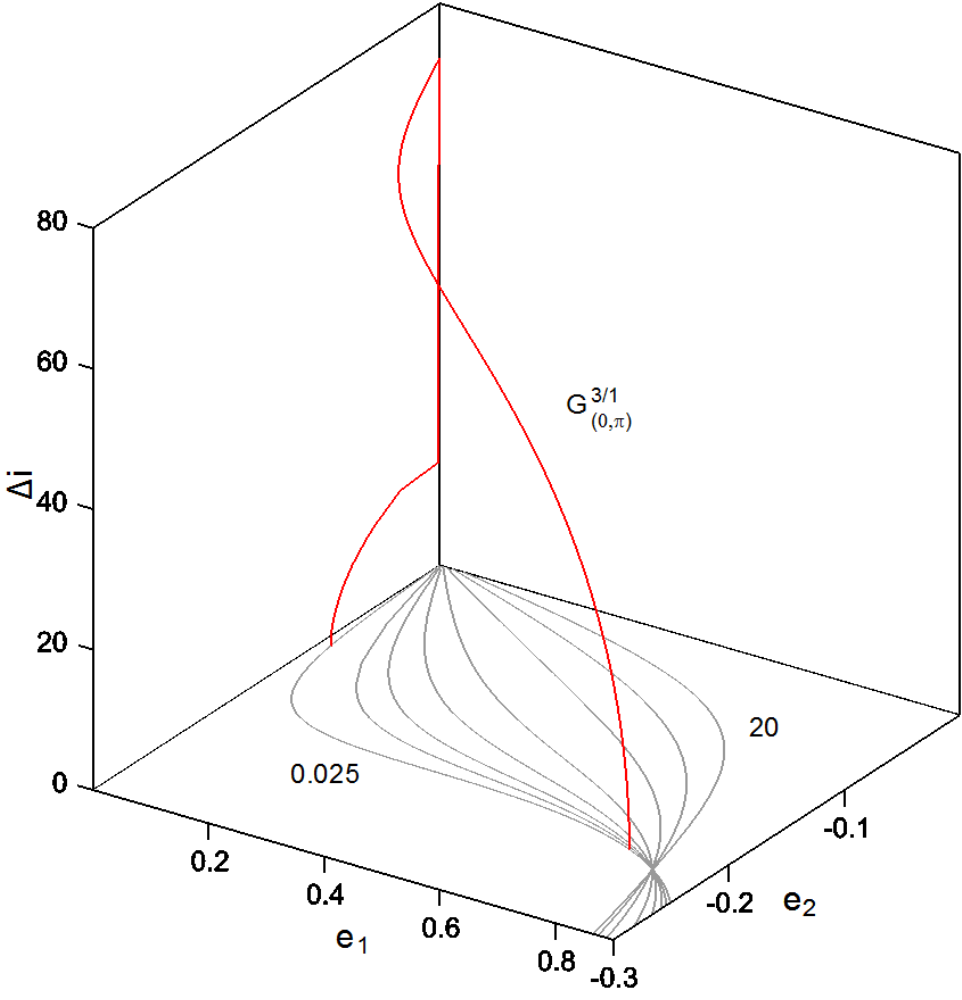}
\end{center}
\caption{ The ${G}^{3/1}_{(0,\pi)}$ family for $\rho=0.025$. The segment for $15^{\circ}<\Delta i<72^{\circ}$ consists of circular orbits.}
\label{31c}
\end{figure}

An interesting characteristic curve is shown for the family ${G}^{3/1}_{(0,\pi)}$ (Fig. \ref{31c}). It starts and ends at the planar family (existence of couple of v.c.o.) and along its way, partially it follows a path with almost circular inclined orbits that extends in the interval $15^\circ<\Delta i<72^\circ$. All orbits along this family are unstable. 
 
\subsection{$4/1$ resonance} 
\begin{figure*}[!htp]
\begin{center}
$\begin{array}{ccc}
\includegraphics[width=6cm,height=6cm]{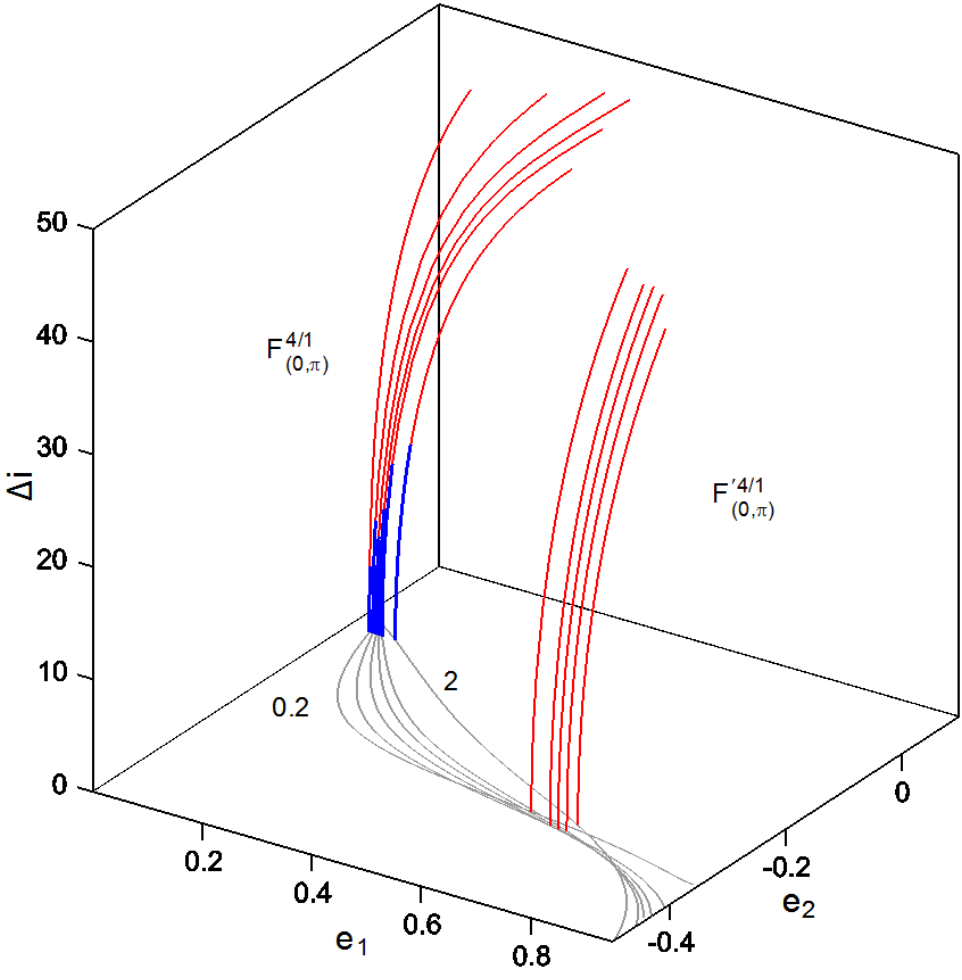}  & \quad&
\includegraphics[width=6cm,height=6cm]{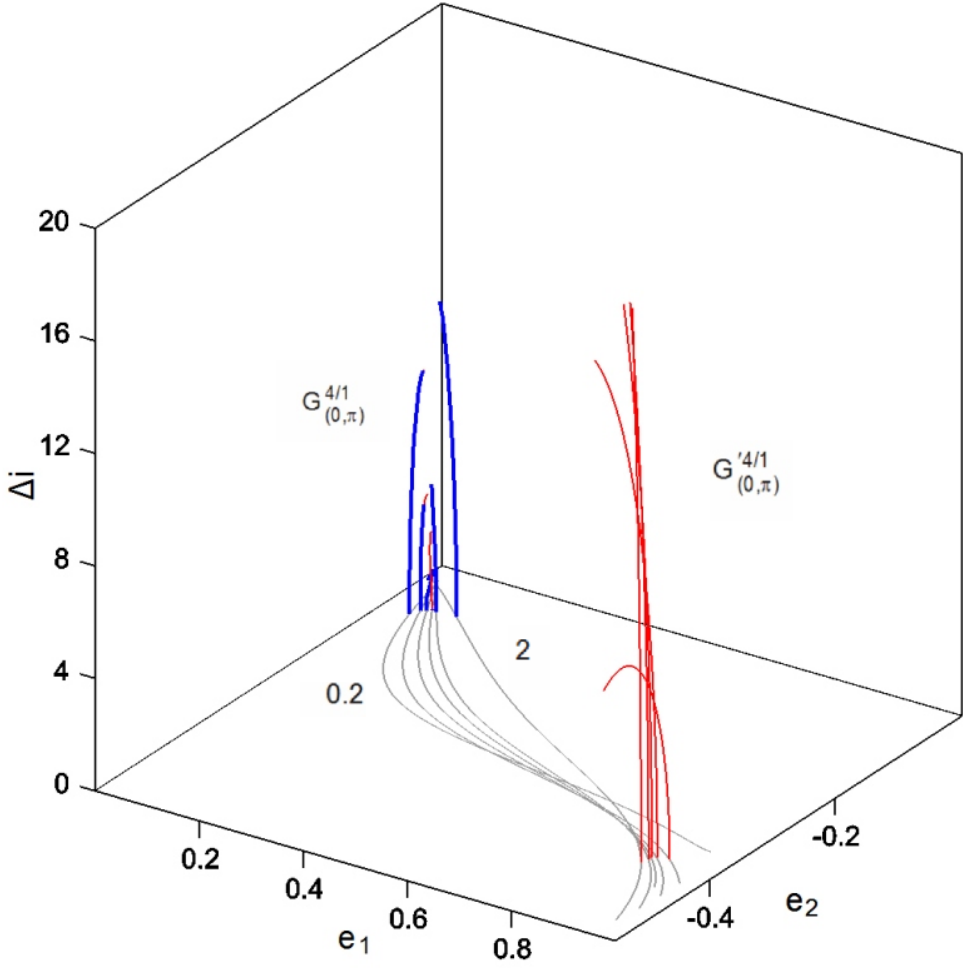} \\
\textnormal{(a)} & \quad & \textnormal{(b)} \\
\includegraphics[width=6cm,height=6cm]{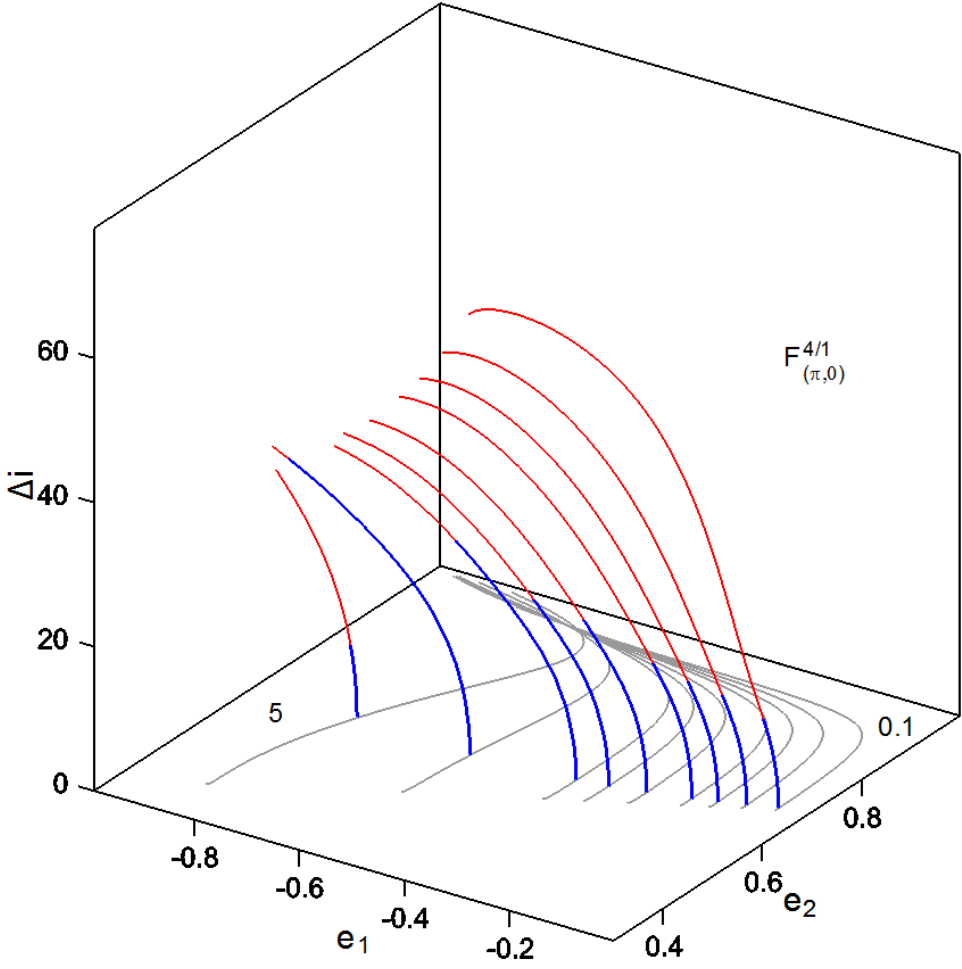}  & \quad&
\includegraphics[width=6cm,height=6cm]{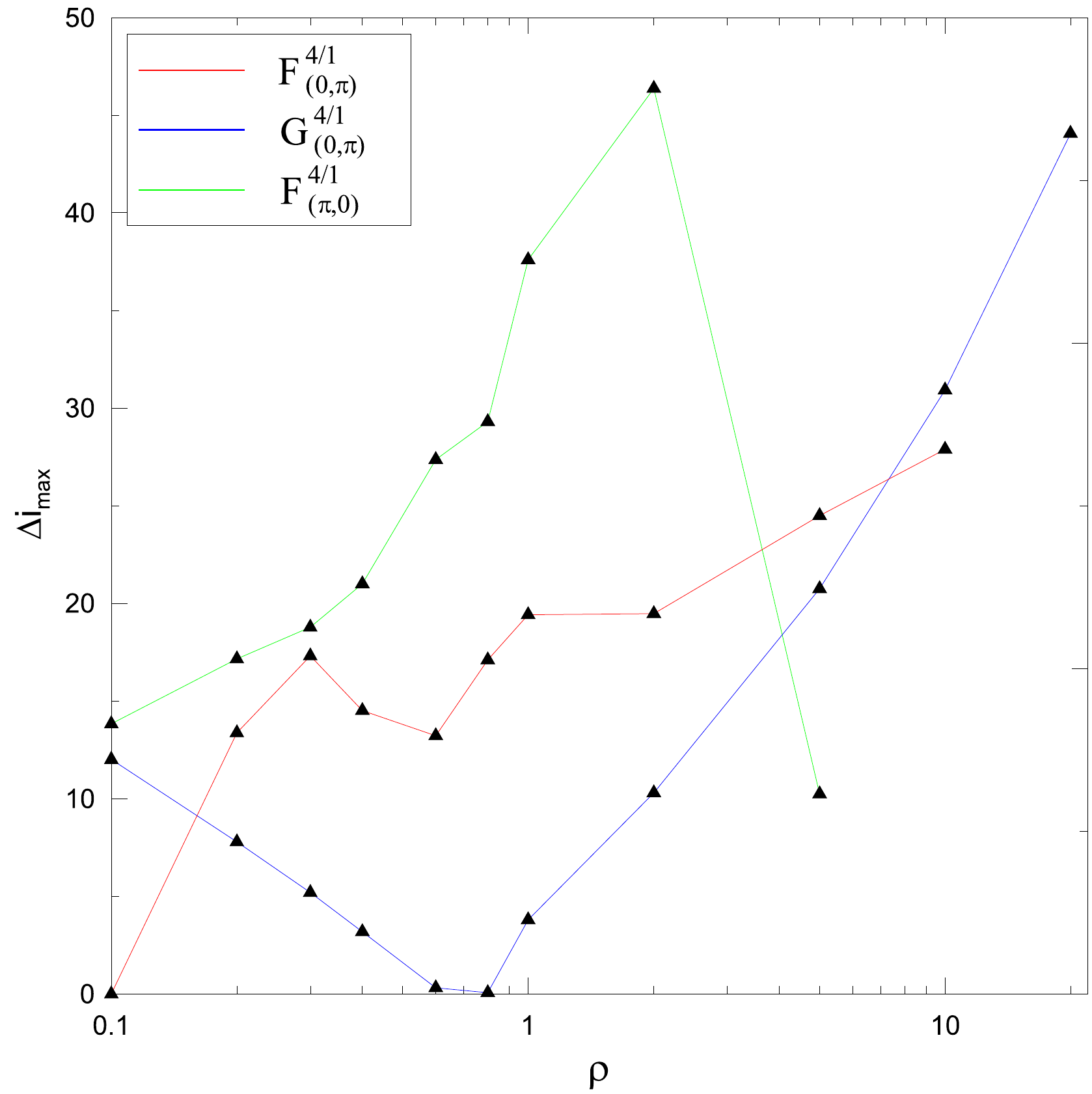} \\
\textnormal{(c)} & \quad &\textnormal{(d)}
\end{array} $
\end{center}
\caption{Spatial families of symmetric periodic orbits 4/1 resonance for various mass ratios. {\bf a} ${F}^{4/1}_{(0,\pi)}$ {\bf b} ${G}^{4/1}_{(0,\pi)}$ and {\bf c} ${F}^{4/1}_{(\pi,0)}$ families. {\bf d} Maximum values of mutual inclination reached along the families as a function of the mass ratio $\rho$.}
\label{41s}
\end{figure*}

In Fig. \ref{41s}{\bf a}, we present the spatial families of $4/1$ resonance ${F}^{4/1}_{(0,\pi)}$ and ${F}^{\prime 4/1}_{(0,\pi)}$. The first ones start having stable orbits for $\rho>0.1$. The stable orbits extend up to a maximum mutual inclination value, which depends on $\rho$, as it is shown in Fig. \ref{41s}{\bf d}( red line). The families ${F}^{\prime 4/1}_{(0,\pi)}$, although some of them ($5 \leq\rho\leq 10$) bifurcate from stable v.c.o., consist of unstable orbits. In Fig. \ref{41s}{\bf b}, the $x$-symmetric families  ${G}^{4/1}_{(0,\pi)}$ and ${G}^{\prime 4/1}_{(0,\pi)}$ are presented. ${G}^{\prime 4/1}_{(0,\pi)}$ are unstable, while families ${G}^{4/1}_{(0,\pi)}$ always start as stable, but in the interval $0.6<\rho<0.8$ stability keeps only for $\Delta i<0.5^\circ$ (Fig. \ref{41s}{\bf d}, blue line). There are also couples of v.c.o. from which the bifurcated families of $x$-symmetric periodic orbits start and end at them forming a bridge. They exist for mass ratios $\rho<0.163$ and are totally unstable; they are not depicted.

In Fig. \ref{41s}{\bf c}, we show the families ${F}^{4/1}_{(\pi,0)}$, which start having stable segments for up to the mass ratio we computed them, i.e. $\rho\leq 5$. Also, in this case we obtain stable orbits of high mutual inclination, up to $47^\circ$ at $\rho=2$. 
We note that the particular v.c.o. correspond to crossing planetary orbits as it happens for any resonance in the configuration $(\pi,0)$.  

\section{Long-term dynamical evolution of orbits}
It is known, that in Hamiltonian systems the stable periodic orbits are surrounded in phase space by invariant tori, where the motion is regular and bounded in the vicinity of them. Therefore, in these regions, as numerical simulations indicate, the long-term stability is guaranteed. In contrary, around unstable periodic orbits in the phase space chaotic regions exists. If chaos is weak, the planetary orbits evolve showing some irregularity in the oscillations of orbital elements, but the configuration of the system does not change significantly for long time intervals. However, in strongly chaotic regions the planetary system is destabilized and the planets show collisions or escapes.   

We hereafter present some numerical results and attempt to connect the above mentioned behaviour of periodic orbits with the long-term dynamical evolution of spatial periodic orbits. We consider initial conditions from an unstable and a stable periodic orbit of the families ${F}^{5/2}_{(0,\pi)}$ for $\rho=0.4$. During numerical integrations the relative error in energy and angular momentum is less than $10^{-11}$ (except in cases of close encounters). The regular or chaotic nature of the system's evolution can be also verified by the F.L.I. mentioned in Sect. \ref{SecStab3D}. 

\subsection{Case i: Long-term evolution of a $3D$ unstable orbit}
We consider the computed (with accuracy $10^{-12}$) initial conditions of an unstable periodic orbit  (from the family mentioned above) that correspond approximately to the orbital elements 
\begin{equation}
\begin{array}{llllll}
a_1=1.61,&e_1=0.27,&i_1=15.96^{\circ},&\Omega_1=90^{\circ},&\omega_1=270^{\circ},&M_1=0^{\circ}\\
a_2=2.97,&e_2=0.55,&i_2=35.83^{\circ},&\Omega_2=270^{\circ},&\omega_2=270^{\circ},&M_2=180^{\circ}
\label{uinit}
\end{array}
\end{equation} 

\begin{figure}[H] 
\begin{center}
$\begin{array}{ccc}
\includegraphics[width=6cm,height=6cm]{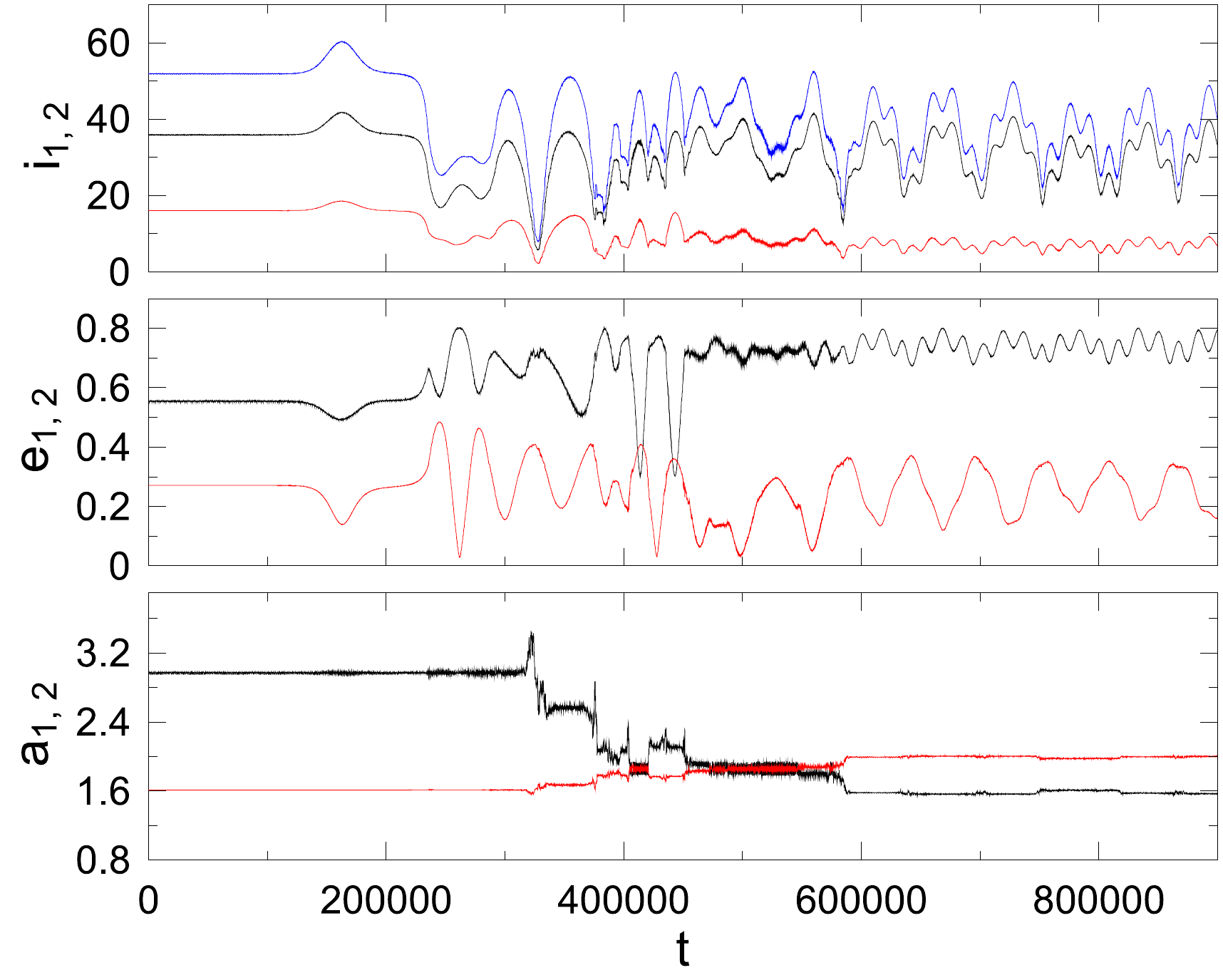}  & \quad&
\includegraphics[width=6cm,height=6cm]{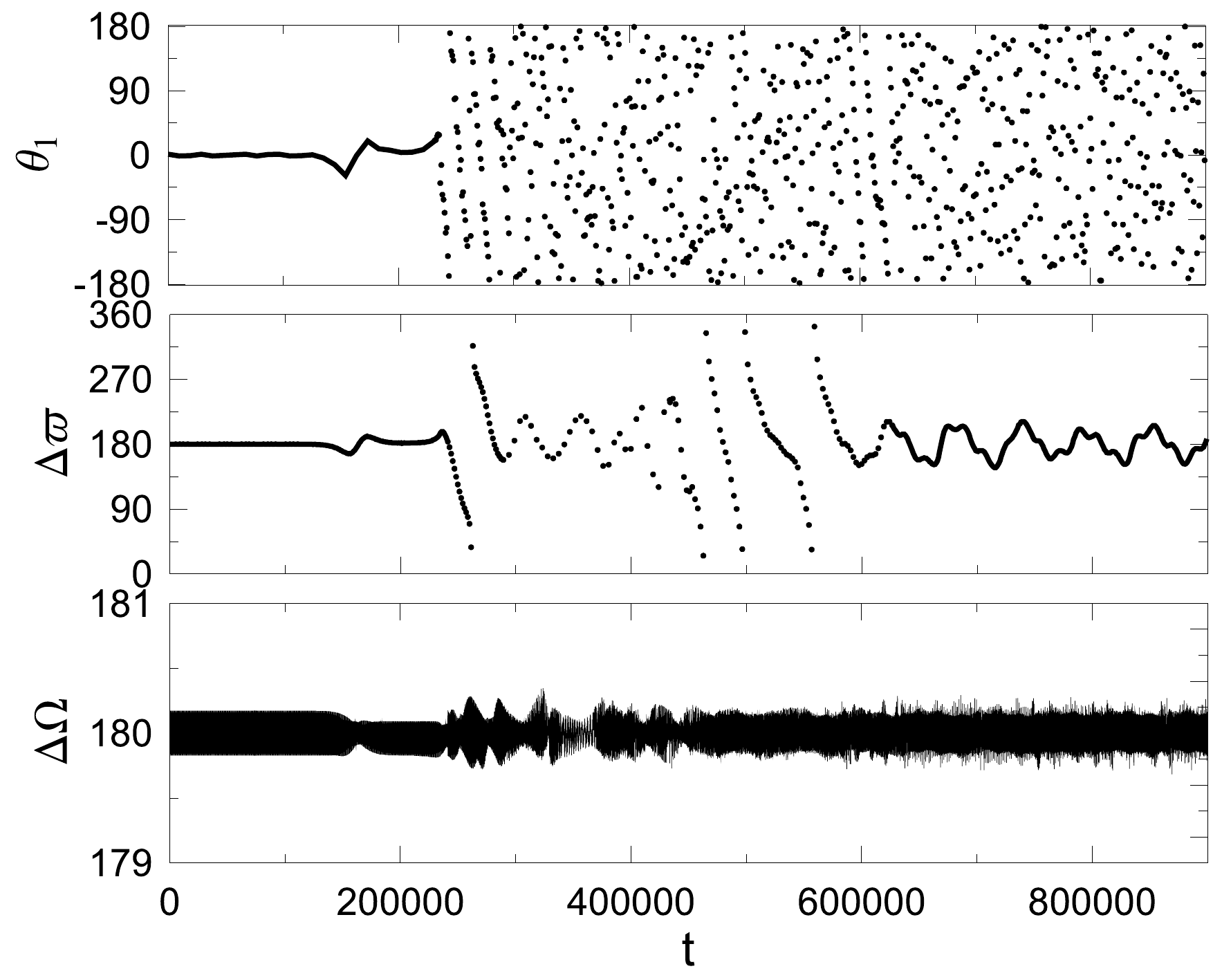} \\
\textnormal{(a)} & \quad & \textnormal{(b)} 
\end{array} $
\end{center}
\caption{The evolution of orbital elements $a_i$, $e_i$, $i_i$, resonant angles $\theta_1$, $\Delta\varpi$ and $\Delta\Omega$ of a planetary system with initial conditions very close to an unstable periodic orbit (\ref{uinit}). Red and black line stands for inner ($P_1$) and outer planet, $P_2$, respectively. Blue line (in the top panel) stands for the mutual inclination $\Delta_i$.}
\label{52uu}
\end{figure}

In Fig. \ref{52uu}, we depict the evolution of the orbital elements $a_i$, $e_i$ and $i_i$ and the resonant angles $\theta_1$, $\Delta\varpi$ and $\Delta\Omega$. Initially, the evolution takes place close to the unstable periodic orbit. The orbital elements remain almost constant and the resonant angles, $\theta_1$, $\Delta\varpi$ and $\Delta\Omega$, librate slightly around $0^\circ$, $180^\circ$ and $180^\circ$, respectively, until $2\times10^{5}$ t.u.. However, due to the instability (and since the initial conditions do not correspond exactly to the periodic orbit), the evolution is drifted along the unstable manifold of the periodic orbit and then strong chaos becomes apparent destabilizing the planetary system. The orbital elements start oscillating irregularly, whereas the resonant angles start to rotate, apart from $\Delta\Omega$, which shows very small libration around $180^\circ$, due to the conservation of the angular momentum. The planets show temporary captures in different resonances and at the end of the evolution are found to be captured in $2/3$ resonance (namely, planet $P_2$ became the inner planet) and $\Delta\varpi$ librates. 
\subsection{Case ii: Long-term evolution of a $3D$ stable orbit}
\begin{figure}[H]
\begin{center}
$\begin{array}{ccc}
\includegraphics[width=6cm,height=6cm]{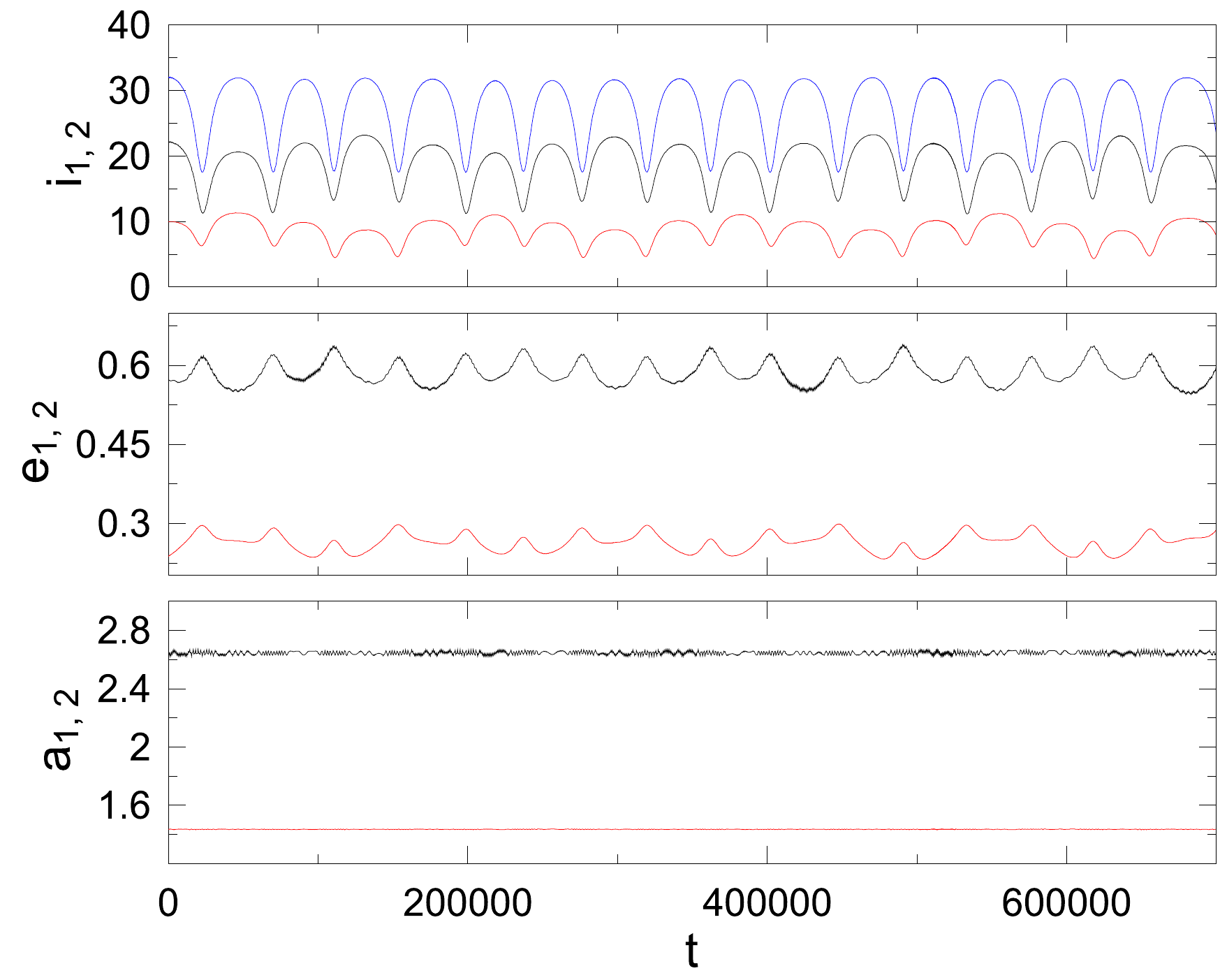} & \quad&
\includegraphics[width=6cm,height=6cm]{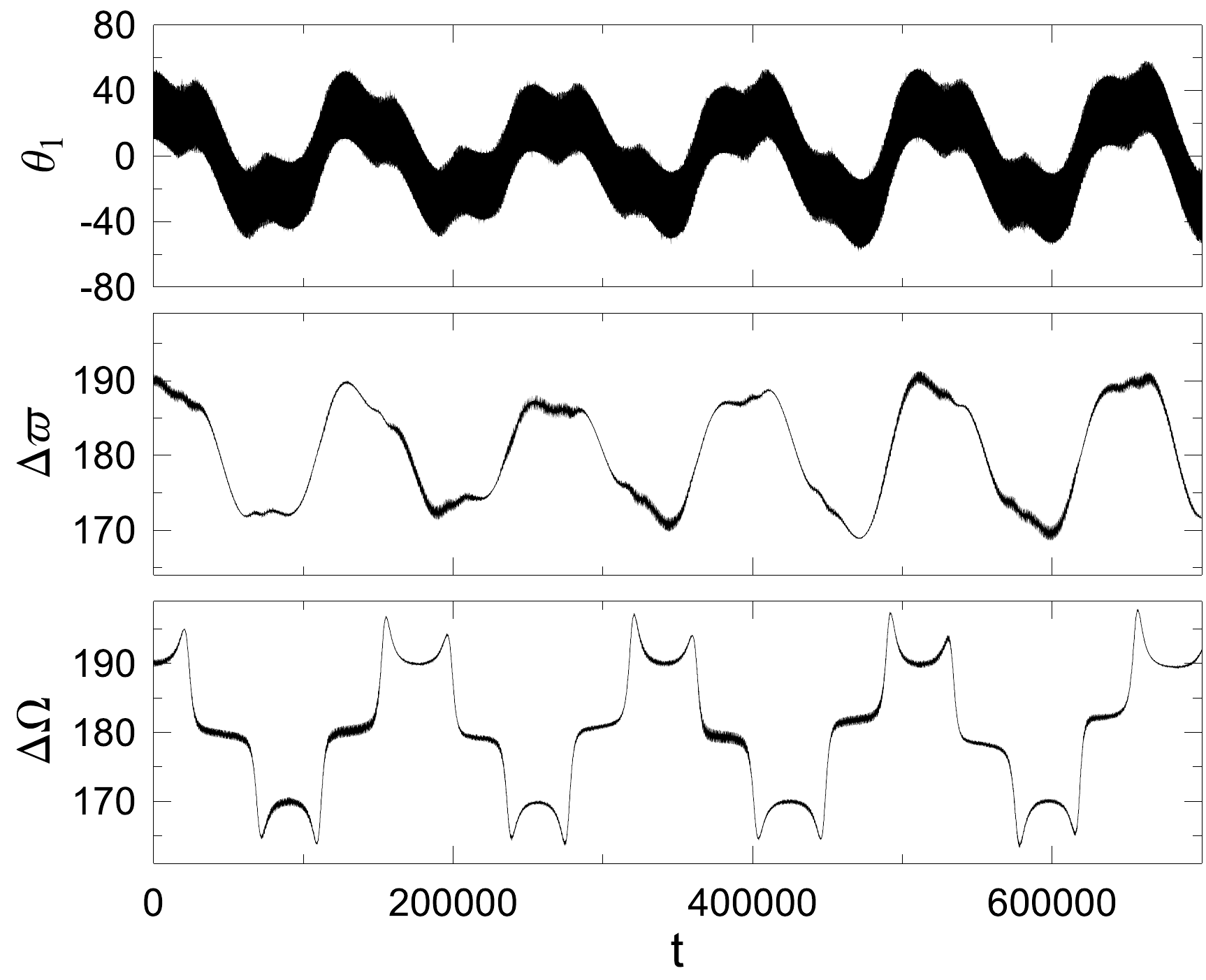} \\
\textnormal{(a)} & \quad & \textnormal{(b)} 
\end{array} $
\end{center}
\caption{Long-term evolution of a planetary system with initial conditions as those of the stable periodic orbit (\ref{sinit}) but with $\Omega_2=280^\circ$ (presentation as in Fig. \ref{52uu}).}
\label{52ss}
\end{figure}

We consider the stable periodic orbit: 
\begin{equation}
\begin{array}{llllll}
a_1=1.43,&e_1=0.23,&i_1=9.94^{\circ},&\Omega_1=90^{\circ},&\omega_1=270^{\circ},&M_1=0^{\circ}\\
a_2=2.64,&e_2=0.56,&i_2=22.13^{\circ}, &\Omega_2=270^{\circ},&\omega_2=270^{\circ},&M_2=180^{\circ}
\label{sinit}
\end{array}
\end{equation}
During the evolution of the exact periodic orbit, the osculating elements and the resonant angles are almost constant and, approximately, equal to their initial values.  However, if we consider orbits with initial conditions that deviate from the initial conditions of the stable periodic orbit, we should obtain quasiperiodic regular evolution. For instance, if we use the initial conditions (\ref{sinit}), but set the initial ascending node of the outer planet $\Omega_2=280^\circ$, we obtain the evolution shown in Fig. \ref{52ss}. All orbital elements and the presented angles oscillate regularly around the initial values of the periodic orbit. So, an island of stability should exist around the orbit (\ref{sinit}).    
 
\begin{figure}[H]
\begin{center}
$\begin{array}{ccc}
\includegraphics[width=6cm,height=6cm]{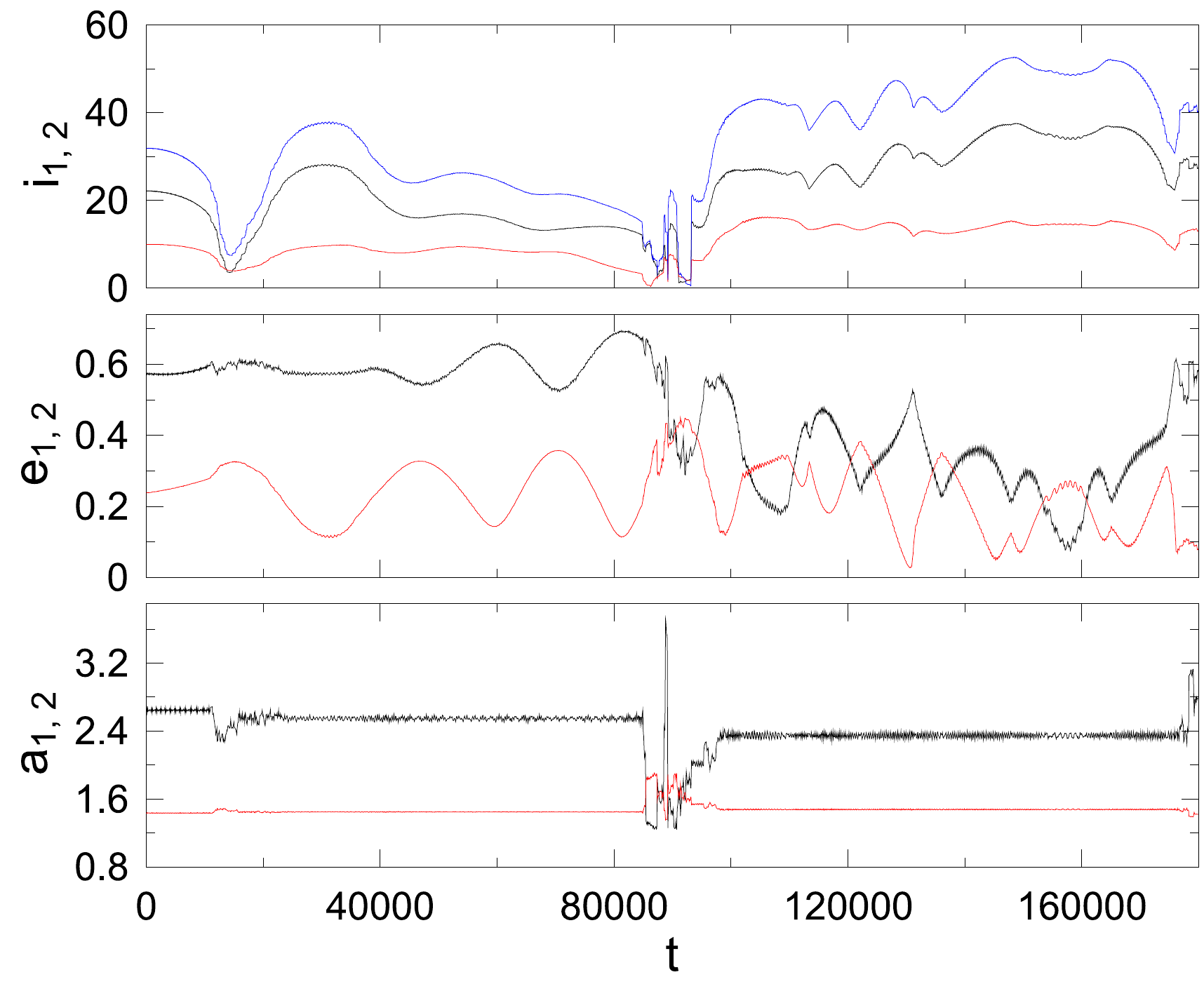}  & \quad&
\includegraphics[width=6cm,height=6cm]{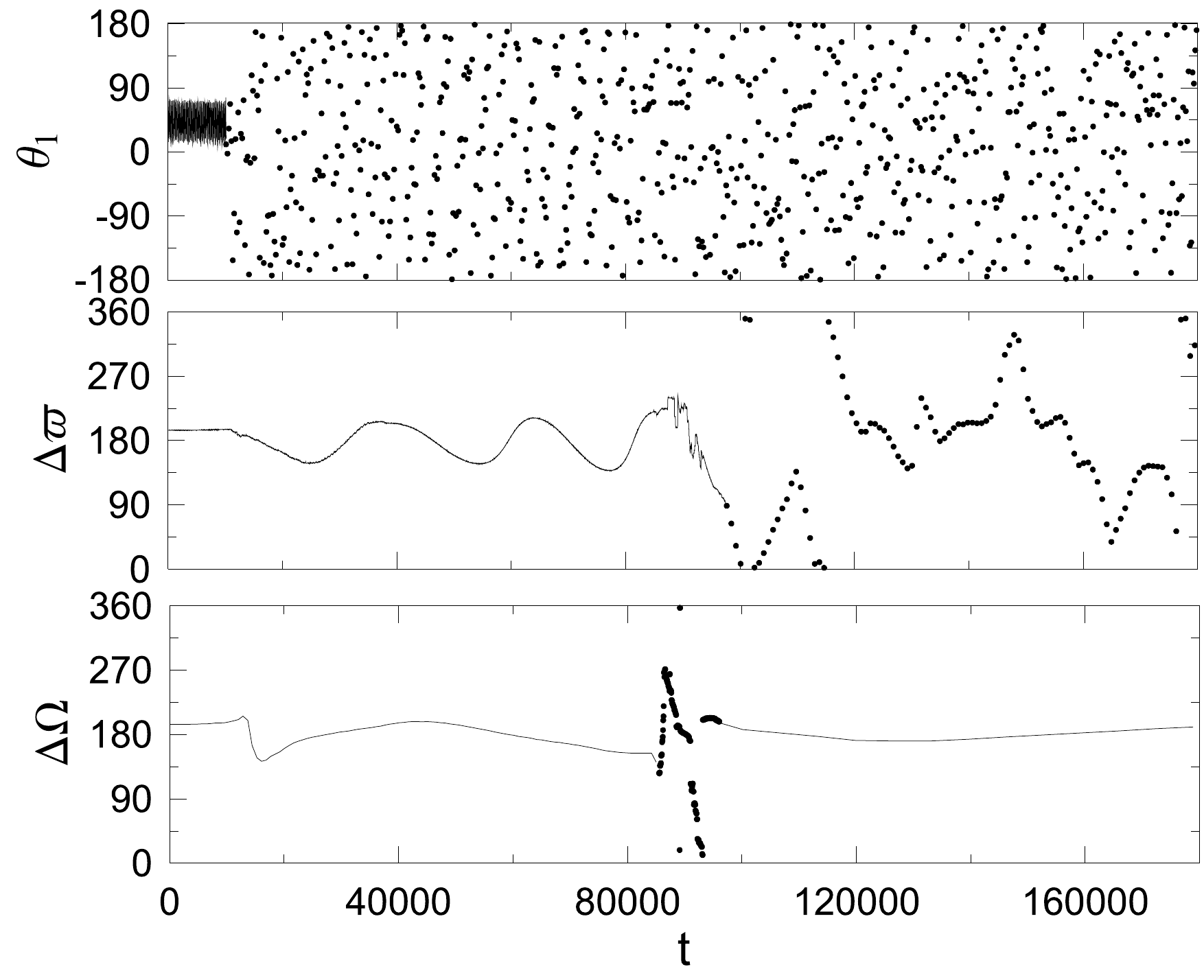} \\ 
\textnormal{(a)} & \quad & \textnormal{(b)} 
\end{array} $
\end{center}
\caption{Long-term evolution of a planetary system with initial conditions as those of the stable periodic orbit (\ref{sinit}) but with $\Omega_2=284^\circ$ (presentation as in Fig. \ref{52uu}).}
\label{52su}
\end{figure}

If we do not start sufficiently close to the stable periodic orbit, i.e. if we increase the initial $\Omega_2$ value by $14^\circ$ with respect to the periodic orbit (\ref{sinit}), we enter a chaotic region in phase space. The evolution is presented in Fig. \ref{52su}. Until $10^4$ t.u. the orbital elements and the presented angles librate around their expected values. Then, although $\Delta\varpi$ continues to librate, $\theta_1$ rotates indicating that the system left the region around the periodic orbit. In the interval $9-9.3\times10^4$ close encounters between the planets take place. Then, slow rotation of the apsidal difference $\Delta\varpi$ is observed besides the rotation of $\theta_1$. Finally, we observe that the planet $P_2$ is scattered.
  
\section{Conclusions}
Using the general three body problem as a model for the study of the dynamics of resonant planetary systems, we determined families of periodic orbits in the planar and spatial case and examined their stability.  The orbital periodicity refers to a particular rotating frame for the model. Such periodic orbits indicate the exact mean motion resonances in phase space. We have performed an extensive study and provide results for $4/3$, $3/2$, $5/2$, $3/1$ and $4/1$ resonances.    

In the planar problem, we presented the families of symmetric periodic orbits in the plane of eccentricities $(e_1,e_2)$ for various values of planetary mass ratio. All symmetric periodic orbits can be classified in four different configurations, which are distinguished by the different values of the resonant angles ($\theta_1,\theta_2)$, namely we have the cases $(0,0)$, $(0,\pi)$, $(\pi,0)$ and $(\pi,\pi)$. In cases like 2nd order resonances the different configurations are distinguished by the pairs ($\theta_3,\theta_1)$.  Some common features for all the examined resonances are the following:
\begin{itemize}
\item The majority of periodic orbits in the configuration $(\pi,\pi)$ is unstable.
\item In the configuration $(\pi,0)$, we always obtain a phase protection mechanism and we have stable planetary orbits that cross each other.
\item In the configuration $(0,\pi)$, we obtain families of stable periodic orbits that bifurcate from almost circular orbits (we remark that close to circular orbits, the configuration may change, however, the generated families consist of stable segments, whose majority of orbits belongs to the $(0,\pi)$ configuration).
\end{itemize}
    
We determined the vertical stability for all computed planar periodic orbits. Orbits that are critical with respect to their vertical stability (vertical critical orbits or, briefly, v.c.o.) are analytically continued to the third dimension and therefore, are bifurcation points of families of periodic orbits in the spatial model. V.c.o. can be found to any configuration, but in our study, no v.c.o. were found in $(\pi,\pi)$ case. Also, they can be stable or unstable with respect to their horizontal stability. From unstable v.c.o., the analytic continuation always generates unstable spatial periodic orbits. From stable v.c.o., either stable, or unstable periodic orbits may bifurcate in space. Nevertheless, our computations showed that in most cases stable v.c.o generate stable families of periodic orbits. Certainly, the stability type can change along the families.        
    
We computed and presented families of spatial periodic orbits for all studied resonances, which are symmetric with respect to either $xz$-plane or $x$-axis. Particularly, we are interested in families starting from stable v.c.o.. Following the analytic continuation of the planar v.c.o. we obtained that stability may extend up to quite large values of planetary mutual inclination (up to $50^\circ$-$60^\circ$ in some cases). The majority of stable spatial periodic orbits is quite eccentric, because the corresponding v.c.o. are quite eccentric, too. An exception occurs for the $4/3$ resonance, where circular inclined orbits have been computed. Also, stable inclined orbits of low inclination values exist in $5/2$ resonance (families $F'^{5/2}_{(0,\pi)}$), but with mutual inclination that does not exceed $12^\circ$.   

Stable periodic orbits are associated with regions in phase space, where the majority of orbits evolves regularly twisting invariant tori. For initial conditions in such regions, a planetary system of two planets shows long-term stability and may, also, survive under small external perturbations (e.g. by additional planets in the system). In contrary, if a planetary system is positioned in the vicinity of an unstable periodic orbit, due to the existence of chaotic regions around it, it will eventually destabilize. This procedure could trigger irregular evolution that generally leads to close encounters between the planets and perhaps forces one planet to be scattered. 

If, indeed, families of periodic orbits constitute paths that can drive the migration process of planets as we mentioned in the introduction, then a planetary system should be found, after resonant capture, in the vicinity of a stable periodic orbit. Therefore, our study can help determine and understand the reasons why, the discovered multi-planet systems possess certain attributes.         

\vspace{1cm}

{\bf Acknowledgments.} This research has been co-financed by the European Union (European Social Fund - ESF) and Greek national funds through the Operational Program ``Education and Lifelong Learning'' of the National Strategic Reference Framework (NSRF) - Research Funding Program: Thales. Investing in knowledge society through the European Social Fund. 

\bibliographystyle{plainnat}
\bibliography{bib}
\end{document}